\documentclass[final]{aa}
\usepackage{txfonts}
\usepackage{graphicx}
\usepackage{color}
\usepackage{multicol}

\usepackage{multirow}
\usepackage{booktabs}   
\usepackage{siunitx} 
\usepackage{hyperref}
\hypersetup{
    colorlinks=true,    
    linkcolor=blue,     
    citecolor=blue,     
    urlcolor=blue       
}

\let\linenumbers\relax
\let\nolinenumbers\relax  

\begin{document}
\title{A Generalist Model Including Evolved Star Mass and Age}
\titlerunning{A Generalist Model Including Evolved Star Mass and Age} 

\author{
    Mengmeng Zhang\inst{1}\
    \and
    Yude Bu\inst{1}\thanks{Corresponding author.}
    \and
    Siqi Wang\inst{1}
    \and
    Shanshan Li\inst{1}
    \and
    Jiangchuan Zhang\inst{2}
    \and
    Jingzhen Sun\inst{1}
    \and
    Yuhang Zhang\inst{1}
    \and
    Ke Wang\inst{1}
    \and
    Jian Liu\inst{3,4,5}
    \and
    Hongliang Yan\inst{6,7}
    \and
    Zhenping Yi\inst{8}
    \and
    Meng Liu\inst{8}
    \and
    Xiaoming Kong\inst{8} 
}
\authorrunning{M. Zhang et al.} 

  \institute{
      School of Mathematics and Statistics, Shandong University, Weihai, 264209, Shandong, China
      \email{buyude@sdu.edu.cn}
      \and
      School of Space Science and Technology, Shandong University, Weihai, 264209, China
      \and
      Weihai Institute for Interdisciplinary Research, Shandong University, Weihai Shandong 264200, China
      \and
      SDU-ANU Joint Science College, Shandong University, Weihai Shandong 264200, China
      \and
      School of Nuclear Science and Technology, University of Science and Technology of China, Hefei 230000, China
      \and
      CAS Key Lab of Optical Astronomy, National Astronomical Observatories, Chinese Academy of Sciences 
      A20 Datun Road, Chaoyang, Beijing 100101, China
      \and
      School of Astronomy and Space Science, University of Chinese Academy of Sciences, Beijing 100049, China
      \and
      School of Airspace Science and Engineering, Shandong University, Weihai, 264209, China
  }  


  \abstract
      {The determination of precise stellar ages and masses for evolved giant stars is fundamental to Galactic archaeology but remains a 
      significant challenge due to the complex degeneracies in stellar spectra. 
      The \textit{Gaia} mission provides low-resolution XP spectra for millions of sources, offering a unique opportunity to infer these 
      evolutionary parameters on a massive scale using data-driven methods.}
      {We aim to extend the capabilities of a transformer-based astronomical foundation model to the domain of evolved stars. 
      Our goal is to establish a unified inference framework that can simultaneously predict standard atmospheric parameters 
      ($T_{\text{eff}}$, $\log g$, $\text{[M/H]}$) and key evolutionary labels (mass and age) from low-resolution spectra, while ensuring 
      physical consistency in the predictions.}
      {Adopting the transformer-based architecture that treats stellar spectra as sequences of tokens, we integrated stellar mass 
      and age into the model's input-output vocabulary. 
      We trained the model using \textit{Gaia} XP spectra cross-matched with the APOGEE DR17 DistMass value-added catalog. 
      Unlike traditional discriminative regressors, our generative approach allows for flexible input handling, including spectral 
      inpainting and parameter-to-spectrum generation.}
      {The model demonstrates high precision on an independent test set, achieving a prediction scatter of 
      $\sigma \approx 0.114 \, M_{\odot}$ for stellar mass and $\sigma \approx 1.334 \, \text{Gyr}$ for age. 
      Beyond numerical accuracy, the model exhibits a deep understanding of stellar physics. 
      It successfully reproduces the non-linear mass-luminosity relation of the giant branch and autonomously disentangles the effects of 
      interstellar extinction from intrinsic temperature variations without explicit physical priors. 
      Furthermore, the model shows robust performance in recovering missing spectral data and estimating reliable uncertainties.}
      {This work validates that foundation models can effectively internalize the physical laws of stellar evolution from data. 
      By providing a physically aware and probabilistic framework for determining stellar ages and masses, this approach offers a powerful 
      tool for disentangling the history of the Milky Way using large-scale spectroscopic surveys.}


   \maketitle
\section{Introduction}\label{sec:intro}
   Understanding the formation and evolution of the Milky Way requires precise characterization of millions of stars across the Galaxy. 
   In the era of large-scale surveys, we have witnessed an explosion of spectroscopic data. 
   Particularly, the \textit{Gaia} mission \citep{2016A&A...595A...1G, 2023A&A...674A...1G} has revolutionized the field by providing 
   low-resolution BP/RP (XP) spectra for hundreds of millions of sources. 
   This massive dataset holds the potential to map the chemo-dynamical history of the Galactic disk in unprecedented detail 
   \citep[e.g.,][]{2002ARA&A..40..487F}. 
   However, fully exploiting this information requires efficient and robust tools capable of extracting physical parameters from complex, 
   high-dimensional spectral data. 
   While ground-based surveys like APOGEE \citep{2017AJ....154...94M} and LAMOST \citep{2012RAA....12.1197C} 
   have provided immense chemical abundance data, the all-sky coverage of \textit{Gaia} offers a unique opportunity for global Galactic 
   archaeology.

   Among the various stellar tracers, evolved giant stars are of particular importance as they serve as standard candles and clocks 
   for probing the Galaxy's distant past \citep{2016ARA&A..54...95G}. 
   However, determining the mass and age of giant stars remains a significant challenge in stellar astrophysics. 
   Unlike atmospheric parameters ($T_{\text{eff}}$, $\log g$, $\text{[M/H]}$), which leave strong imprints on stellar spectra, mass and age 
   are governed by subtle evolutionary states. 
   Traditionally, these parameters are derived via isochrone fitting or asteroseismology. While asteroseismology yields precise ages, it is 
   limited to a small fraction of stars with high-cadence photometry. 
   Consequently, determining ages for the vast majority of giants observed by \textit{Gaia} remains a bottleneck, often hampered by the strong 
   degeneracies between the Red Giant Branch (RGB) and the Red Clump (RC).

   In recent years, data-driven methods, particularly machine learning (ML), have become standard for deriving stellar labels 
   \citep[e.g., \textit{The Cannon},][; \texttt{astroNN}, \citealt{2019MNRAS.483.3255L}]{2015ApJ...808...16N}. 
   While these discriminative models excel at mapping spectra to atmospheric parameters, they often face limitations when applied to 
   evolutionary parameters like mass and age \citep[e.g.,][]{2016MNRAS.456.3655M, 2019MNRAS.484..294D}. 
   Most traditional architectures (such as Random Forests or CNNs; e.g., \citealt{2018MNRAS.475.2978F}) treat spectra as rigid, fixed-length 
   vectors and lack the flexibility to handle missing data or varying noise levels. 
   More recently, advanced architectures have been introduced to tackle these complexities. For instance, the Spectral Transformer 
   \citep[SPT;][]{2024A&A...683A.163Z} successfully applied attention mechanisms to improve the estimation of age and mass for red giant stars. 
   However, even these advanced models typically operate in a discriminative manner---mapping inputs directly to labels---which often limits 
   their ability to capture the full probabilistic manifold of the data or to perform generative tasks like spectral inpainting.

   To overcome these limitations, the concept of ``Foundation Models'' offers a new paradigm. The advent of the Transformer architecture 
   \citep{2017arXiv170603762V}, by treating stellar spectra as sequences of tokens---analogous to words in natural language processing---allows 
   models to learn the intrinsic manifold of stellar physics from vast unlabeled or labeled datasets. 
   Recently, \citet{2024MNRAS.527.1494L} demonstrated the power of this approach by developing a Transformer-based astronomical foundation model. 
   Their work showed that such a model could robustly predict atmospheric parameters and handle missing data through a unified inference 
   framework. 
   However, their study focused primarily on fundamental atmospheric labels, leaving the more challenging domain of evolutionary 
   parameters---mass and age---largely unexplored within this generative framework.

   In this work, we extend the foundation model framework to address the challenge of characterizing evolved giant stars. 
   By combining the generative capabilities of the Transformer architecture with the high-precision evolutionary labels (mass and age) 
   from the APOGEE DR17 DistMass value-added catalog \citep{2024AJ....167...73S}, we train a model capable of simultaneously inferring 
   atmospheric properties and evolutionary states from \textit{Gaia} XP spectra. 
   We demonstrate that this approach not only achieves high prediction precision but also internalizes the underlying physics of stellar 
   evolution, such as the mass-luminosity relation and the disentanglement of interstellar extinction. 
   This represents a significant step towards a universal, physically-aware model for Galactic archaeology.

   The paper is organized as follows. We begin in Section~\ref{sec:data} by compiling the training data set that links Gaia XP 
   observations with ground-truth evolutionary parameters. 
   Section~\ref{sec:methods} presents the extended Transformer architecture and our training strategy. 
   The core results are analyzed in Section~\ref{sec:result}, where we perform bidirectional inference tests mapping spectra to 
   labels and vice versa—to verify the model's robustness and physical consistency. 
   Section~\ref{sec:Discussion} provides a deeper discussion on the learned embedding space and the interpretability of the 
   attention maps. 
   Finally, Section~\ref{sec:conclusion} summarizes the key contributions of this work.

\section{Data preparation}\label{sec:data}
   To train our foundation model, we construct a heterogeneous data set that combines high-resolution spectroscopy, 
   multi-band photometry, astrometry, and derived stellar evolution labels. Our data compilation strategy largely follows 
   that of \citet{2024MNRAS.527.1494L}, utilizing data from APOGEE, \textit{Gaia}, and three-dimensional dust maps. 
   In this work, we significantly extend the utility of the training set by incorporating stellar mass and age estimates, 
   allowing the model to learn the underlying physics of stellar evolution.
   
   \subsection{Base Observational Data} \label{Base Observational Data}
      Our primary spectroscopic labels ($T_{\text{eff}}$, $\log g$, and $[\text{M/H}]$) are drawn from the Apache Point 
      Observatory Galactic Evolution Experiment Data Release 17 \citep[APOGEE DR17;][]{2017AJ....154...28B, 2022ApJS..259...35A}. 
      We utilize the parameters derived from the APOGEE Stellar Parameter and Chemical Abundances Pipeline 
      \citep[ASPCAP;][]{2016AJ....151..144G}. 
      To ensure data quality, we filter out obvious binary stars and problematic measurements using the \texttt{STARFLAG} and 
      \texttt{ASPCAPFLAG} bitmasks, treating flagged parameters as missing data (NaN) rather than discarding the stars entirely.

      Astrometric and spectrophotometric data are obtained from \textit{Gaia} Data Release 3 \citep[\textit{Gaia} DR3;][]{2016A&A...595A...1G, 2023A&A...674A...1G}. 
      We employ the low-resolution BP/RP (XP) spectra, which are represented as coefficients of an orthogonal basis function 
      expansion. 
      Following the methodology of \citet{2024MNRAS.527.1494L}, we normalize these coefficients by the \textit{Gaia} $G$-band 
      flux to remove the dependence on distance modulus (see their Eq. 8). 
      We adopt the parallax zero-point corrections from \citet{2023A&A...674A...2D} and \citet{2023MNRAS.519..948L}, and we 
      calculate a "pseudo-luminosity" $L_{\text{pseudo}}$ (the inverse of the parallax squared, scaled by apparent magnitude) to serve as 
      a luminosity proxy that preserves the Gaussianity of parallax errors.

      Interstellar extinction estimates are retrieved from the "Combined19" three-dimensional dust map using the 
      \texttt{mwdust} package \citep{2016ApJ...818..130B}. 
      This map combines information from \citet{2003A&A...409..205D}, \citet{2006A&A...453..635M}, and \citet{2019ApJ...887...93G}. 
      We use the median value of the logarithmic reddening, $\ln E(B-V)$, derived from 1000 Monte Carlo samples of 
      the distance modulus distribution.
      \begin{figure*}
         \begin{center}
            \includegraphics[width=\linewidth]{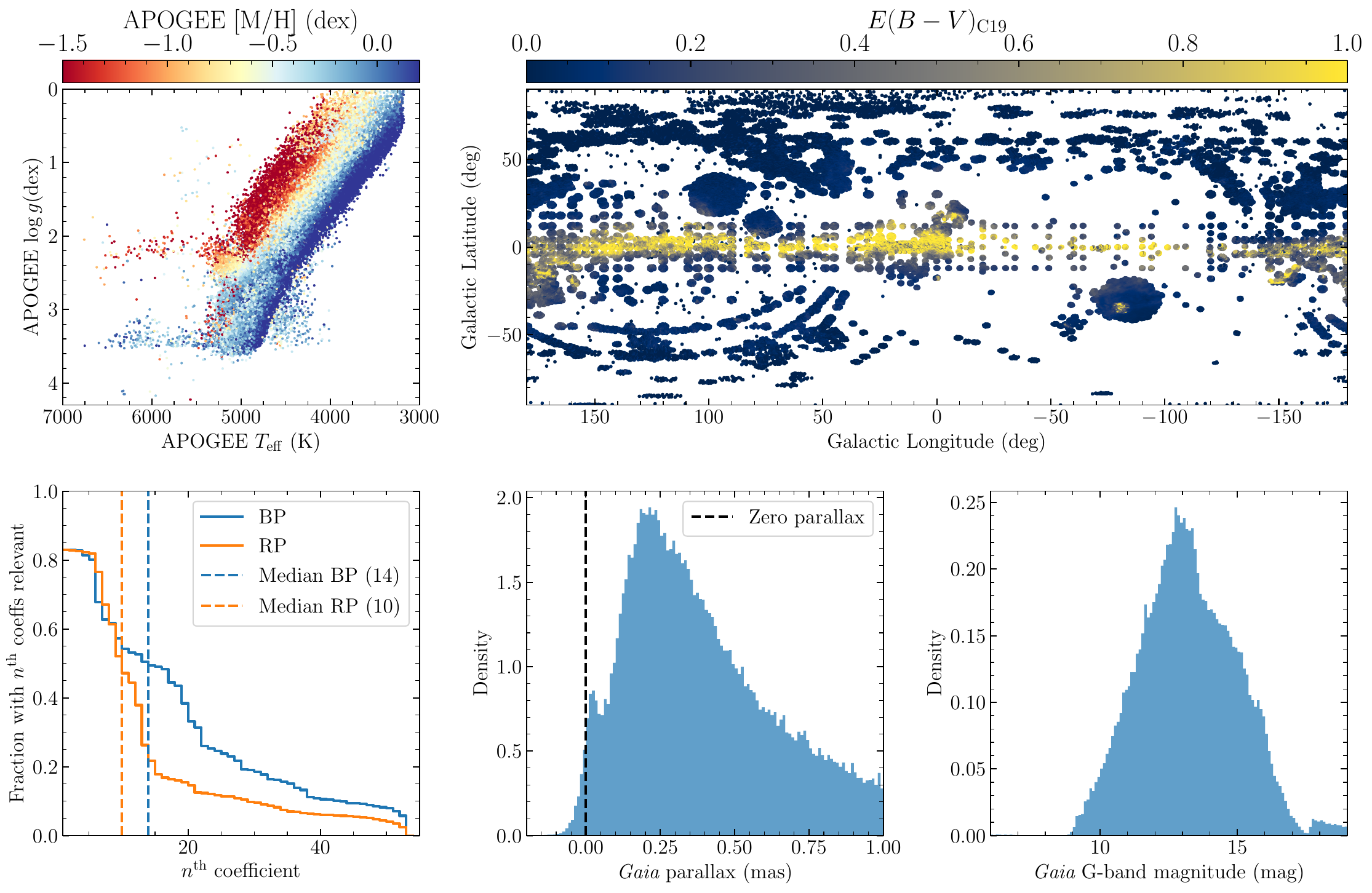}  
            \end{center}
            \caption{Properties of the training data set constructed for the foundation model.  
               \textit{Top left}: The Kiel diagram ($T_{\text{eff}}$ vs. $\log g$) for sources with valid APOGEE labels, 
               color-coded by $\mathrm{[M/H]}$.
               The sample is dominated by giants. 
               \textit{Top right}: Sky distribution of the training sources in Galactic coordinates $(l, b)$, color-coded by the 
               interstellar extinction $E(B-V)$ from the Combined19 map. The distribution traces the APOGEE survey footprint. 
               \textit{Bottom left}: The fraction of stars for which the $n$-th \textit{Gaia} XP coefficient is marked as `relevant' by 
               the \textit{Gaia} pipeline. 
               The vertical dashed lines indicate the median number of relevant coefficients for BP (blue) and RP (orange). 
               \textit{Bottom right}: Distributions of \textit{Gaia} parallax (left sub-panel) and $G$-band magnitude (right sub-panel) 
               for the training sample.}
         \label{fig:training_set_foundation}
      \end{figure*}

      \begin{figure}[!h]
         \begin{center}
         \includegraphics[width=\linewidth]{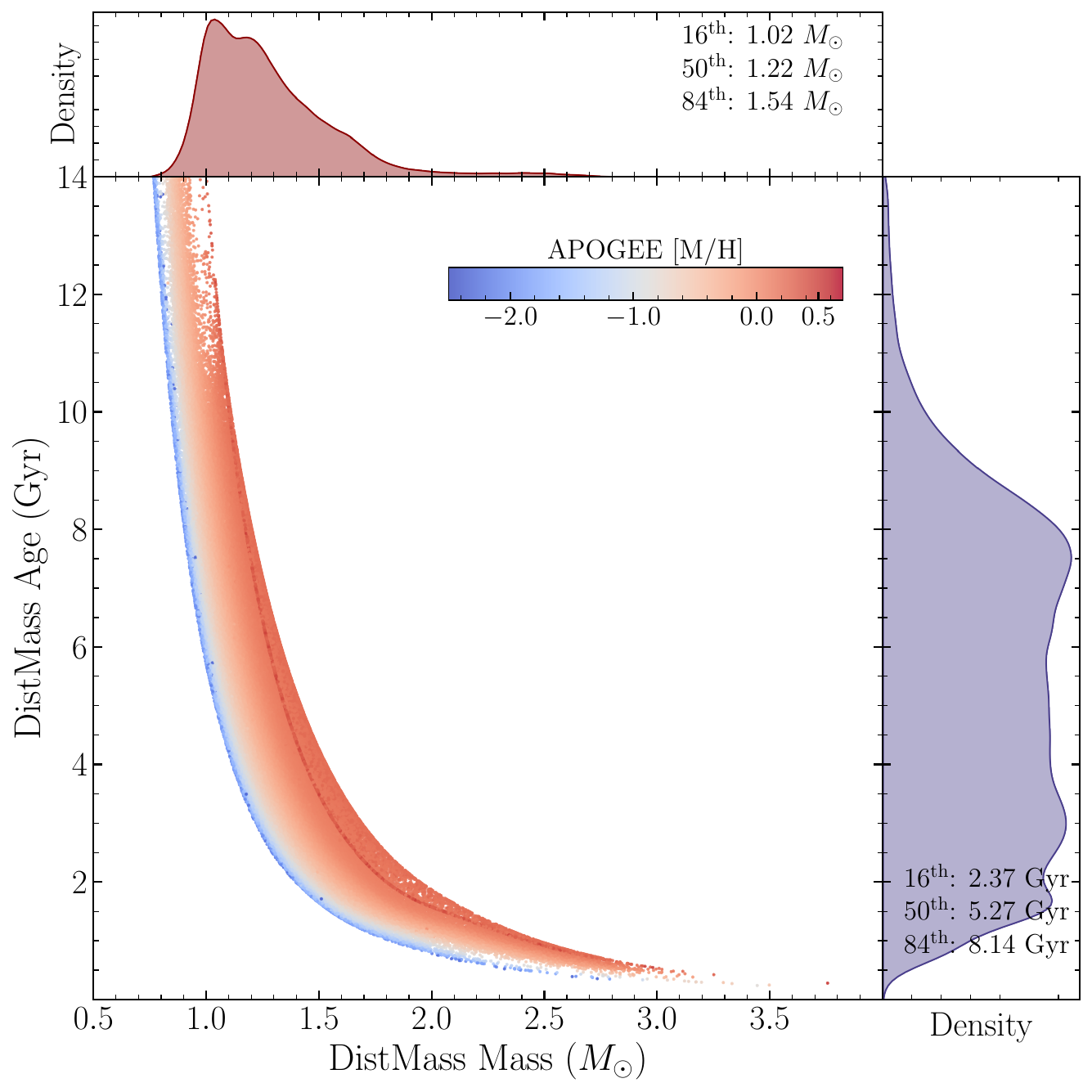}  
         \end{center}
         \caption{
            The joint distribution of stellar mass, age, and $[\mathrm{M/H}]$ for the training sample after quality cuts. 
            \textit{Main panel}: Scatter plot of stellar age versus mass, color-coded by APOGEE $[\mathrm{M/H}]$. 
            The correlation between mass, age, and $[\mathrm{M/H}]$ reflects the expected Galactic chemical evolution trends. 
            \textit{Top panel}: The marginal kernel density estimation (KDE) of stellar mass. 
            The distribution is strongly peaked around $\sim 1.1\,M_{\odot}$, corresponding to the dominant population of RC stars. 
            \textit{Right panel}: The marginal KDE of stellar age, showing a broad coverage of the Galactic star formation history 
            from young ($<1\,\mathrm{Gyr}$) to old ($>10\,\mathrm{Gyr}$) populations.}
         \label{fig:mass_age_metallicity_distribution}
      \end{figure}

   \subsection{Stellar mass and age Labels} \label{Stellar Mass and Age Labels}
      To provide ground truth labels for stellar mass ($M$) and age ($\tau$), we cross-match our base catalog with the value-added 
      catalog of \citet{2024AJ....167...73S}. 
      This catalog provides mass and age estimates for APOGEE DR17 stars derived via isochrone matching techniques. 
      These physical parameters serve as the key training targets that enable our model to infer evolutionary stages.

   \subsection{Data preprocessing and vector construction} \label{Data Preprocessing and Vector Construction}
      We integrate the observational data and physical labels into a unified 120-dimensional input vector for each star. 
      This vector comprises: 
      (1) the 110 normalized \textit{Gaia} XP coefficients; 
      (2) broad-band colors $(G_{\text{BP}}-G_{\text{RP}})$, $(J-H)$, and $(J-K_s)$; 
      (3) APOGEE atmospheric parameters; 
      (4) extinction and pseudo-luminosity; 
      and (5) the stellar mass and age labels.

      Special attention is paid to the \textit{Gaia} XP coefficients. 
      The number of "relevant" coefficients for a given star is determined by the \textit{Gaia} pipeline based on significance 
      thresholds. 
      To handle the edge cases where the relevancy flags for the highest-order coefficients 
      (indices 52--54 for BP and 107--109 for RP) are occasionally missing, we propagate the relevancy flag from the preceding 
      index (51 and 106, respectively) to these final coefficients. 
      Coefficients deemed irrelevant are set to NaN to ensure the model processes only significant spectral features.

      We construct a corresponding error vector of the same dimension. 
      To maintain numerical stability during training, any missing values (NaNs) in the error vector are replaced with zeros (0.0), 
      while missing values in the input vector remain as NaNs to be handled by the model's masking mechanism.

   \subsection{Sample selection and statistics} \label{Sample Selection and Statistics}
      We perform a cross-match between the base APOGEE-\textit{Gaia} dataset and the mass-age reference catalog. 
      To ensure the physical consistency of the training set, we apply the following quality cuts to the derived labels:
      \begin{enumerate}

      \item \textbf{Mass:} $0.5 < M/M_{\odot} < 4.0$, covering the range from low-mass red giants to intermediate-mass stars.

      \item \textbf{Age:} $0 < \tau < 14$ Gyr, respecting the cosmic age limit.

      \item \textbf{Metallicity:} $-2.5 < [\text{M/H}] < 0.7$, encompassing both halo and disk populations.

      \end{enumerate}

      The final curated data set contains a total of 319,585 stars. 
      We randomly split this sample into a training set of 287,626 stars ($\sim$90\%) and a testing set of 31,959 stars ($\sim$10\%). 
      It is important to note that due to the heterogeneous nature of our dataset, not all stars possess every label; 
      for instance, approximately 74\% (213,829) of the training stars have valid spectroscopic parameters from APOGEE, while the 
      remainder rely on \textit{Gaia} spectrophotometry.

      Figure \ref{fig:training_set_foundation} illustrates the spatial and physical distribution of the training sample. 
      The sample traces the APOGEE survey footprint and is dominated by stars on the RGB and the RC, as evidenced by the Kiel diagram 
      ($T_{\text{eff}}$ vs. $\log g$). 
      The distribution of \textit{Gaia} parallaxes is peaked near zero with a positive tail, consistent with a sample dominated by 
      distant giant stars.

      In Figure \ref{fig:mass_age_metallicity_distribution}, we present the joint distribution of the training labels. 
      The mass distribution (top panel) exhibits a prominent peak at $M \approx 1.2 M_{\odot}$, characteristic of the RC population, 
      which serves as a standard candle. 
      The age distribution (right panel) shows a broad coverage of the Galactic star formation history. 
      The main panel highlights the expected correlation between mass, age, and $[\mathrm{M/H}]$, where massive stars are predominantly 
      young and metal-rich, validating the physical integrity of our training labels.
   \begin{figure*}
      \begin{center}
         \includegraphics[width=\linewidth]{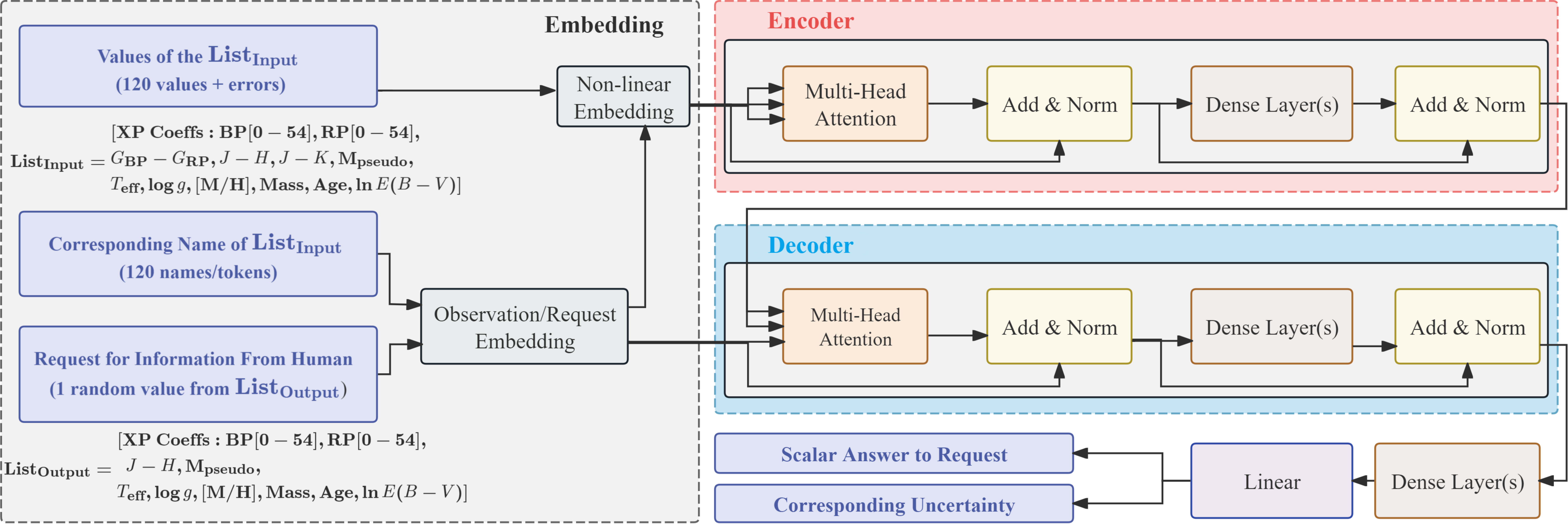}  
         \end{center}
         \caption{Schematic architecture of the Transformer-based foundation model tailored for evolved giants. 
            \textit{Left}: Embedding Module. The model ingests a heterogeneous list of inputs, comprising: 
               (1) \textit{Gaia} XP spectroscopic coefficients; 
               (2) stellar colors (e.g., $G_{BP}-G_{RP}$, $J-H$); 
               (3) atmospheric parameters ($T_{\text{eff}}$, $\log g$, $[\mathrm{M/H}]$); 
               and (4) evolutionary parameters (explicitly including mass and age). 
               Values and their corresponding token names are mapped into a latent space via a non-linear embedding layer. 
            \textit{Top Right}: Encoder. It processes the sequence of embedded observations using multi-head self-attention to 
            capture the dependencies between spectra, atmospheric parameters, and evolutionary labels, generating a context 
            vector representing the star. 
            \textit{Bottom Right}: Decoder. It receives a specific ``information request'' token (e.g., querying ``mass'') and 
            queries the encoder's context via cross-attention. 
            The output yields a probabilistic prediction, consisting of a scalar mean value and its associated uncertainty.}
      \label{fig:RG_model}
   \end{figure*}

\section{Methods}\label{sec:methods}
   In this work, we adopt and extend the Transformer-based ``foundation model'' architecture originally developed by 
   \citet{2024MNRAS.527.1494L}. 
   Unlike traditional architectures that require fixed-dimensional inputs, this data-driven framework treats astronomical 
   observations as sequences of tokens, allowing for the flexible handling of heterogeneous and missing data. 
   We specifically tailor this architecture to infer stellar evolutionary parameters by incorporating stellar mass and age 
   directly into the model's embedding space.

   \subsection{Transformer-based encoder-decoder architecture} \label{Transformer-based encoder-decoder architecture}
      Our model utilizes an asymmetric encoder-decoder structure based on the Transformer architecture \citep{2017arXiv170603762V}. 
      The workflow is illustrated in Figure~\ref{fig:RG_model}.

      \textbf{Encoder:} The encoder ingests a sequence of observed data tokens. 
      It consists of two stacked Transformer blocks separated by dense layers with 1024 and 512 units, respectively. 
      Through multi-head self-attention mechanisms (with 16 heads), the encoder captures the inter-dependencies among the input 
      tokens and compresses the information into a latent context vector that represents the physical state of the star.

      \textbf{Decoder:} The decoder generates predictions based on the encoder's context vector and a specific ``query vector.'' 
      It comprises three Transformer blocks with intermediate dense layers of 3096, 1548, and 774 units. 
      The decoder utilizes cross-attention to map the stellar context to the desired output physical quantity (e.g., age or mass).

      \textbf{Probabilistic Output:} The final stage consists of a three-layer dense network that maps the decoder's output to 
      a probabilistic estimate. 
      For any requested parameter, the model predicts both a value and a corresponding uncertainty, enabling the quantification 
      of aleatoric uncertainty.
      \begin{figure*}
            \begin{center}
               \includegraphics[width=\linewidth]{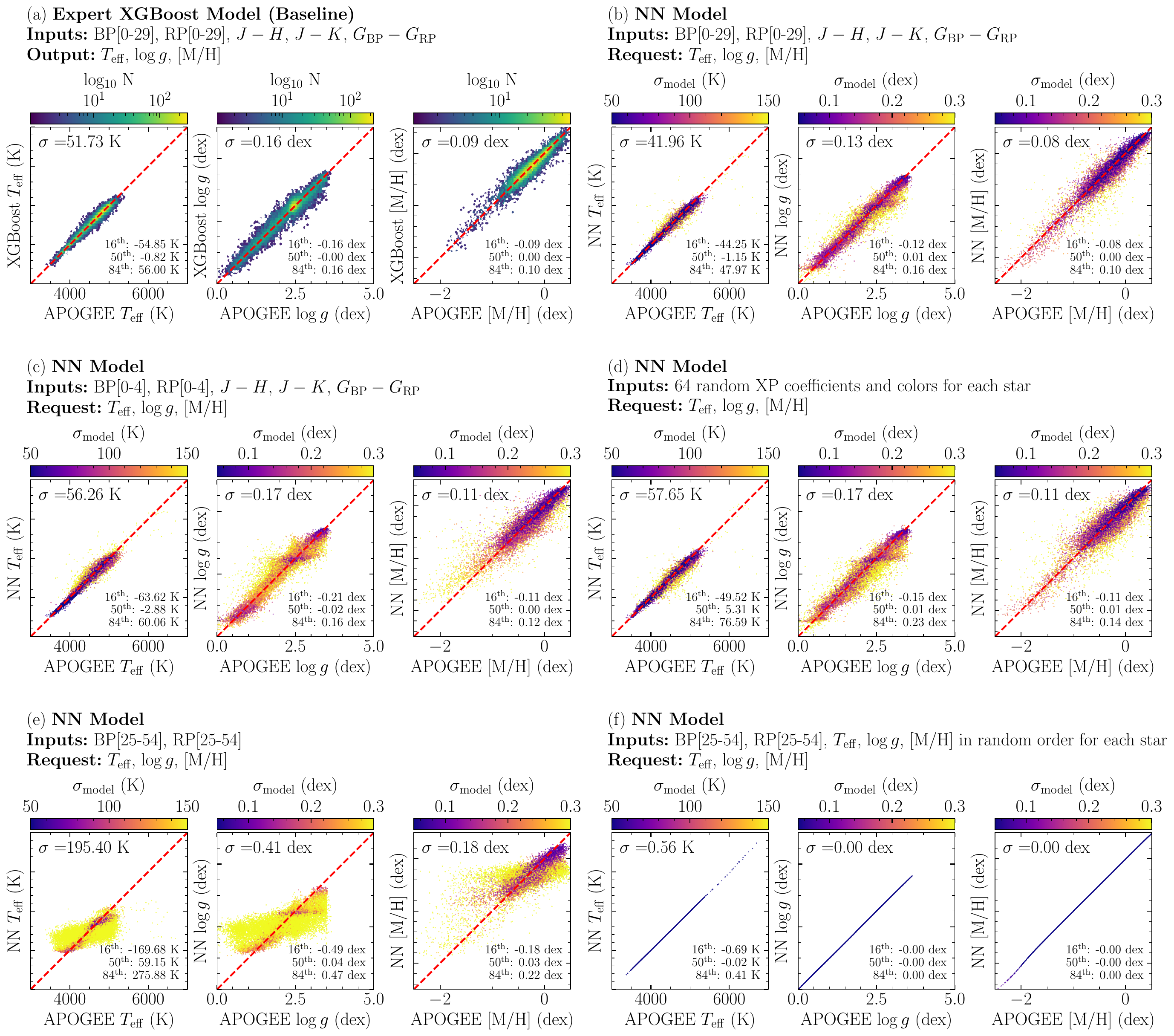}  
               \end{center}
               \caption{
                  Results for predicting stellar atmospheric parameters ($T_{\text{eff}}$, $\log g$, and $\text{[M/H]}$) from 
                  different combinations of XP spectra and photometry. 
                  \textit{(a)} An expertly trained XGBoost baseline model using standard inputs. 
                  \textit{(b)} Our Transformer-based model with the same inputs, showing improved precision. 
                  \textit{(c)} Performance using only the first 5 BP/RP coefficients. 
                  \textit{(d)} Performance using a random subset of 64 XP coefficients and colors. 
                  \textit{(e)} A stress test using uninformative high-order coefficients, where the model correctly predicts large 
                  uncertainties. 
                  \textit{(f)} An identity recovery test where ground truth labels are mixed into the inputs.}
            \label{fig:spectra_to_params_atm}
         \end{figure*}

   \subsection{Embedding space and parameter integration} \label{Embedding space and parameter integration}
      A critical modification in this study is the explicit integration of stellar mass and age into the model's 
      token vocabulary. 
      To process continuous physical values (fluxes, atmospheric parameters, mass, and age) within a Transformer architecture, 
      we employ the non-linear embedding technique described in \citet{2024MNRAS.527.1494L}.

      For any given observation type $x$ (which can be a specific \textit{Gaia} XP coefficient, $T_{\text{eff}}$, $M$, or $\tau$) 
      with a scalar value $v_x$, we map it to a $d_{\text{model}}=128$ dimensional embedding vector $\mathbf{e}_x$:
      \begin{equation}
         \mathbf{e}_x = \mathrm{GeLU}(\mathbf{w}_x \cdot v_x + \mathbf{b}_x)
         \label{eq:embedding}
      \end{equation}
      where $\mathbf{w}_x$ and $\mathbf{b}_x$ are trainable weight and bias vectors unique to the data type $x$, and GeLU is the 
      Gaussian Error Linear Unit activation function. 
      By defining $M$ and $\tau$ as independent tokens with their own learnable embedding weights ($\mathbf{w}_M$, $\mathbf{w}_\tau$), 
      these evolutionary parameters acquire the same status as spectroscopic data in the latent space. 
      This allows the model to function bidirectionally: estimating mass/age from spectra, or using mass/age constraints to 
      reconstruct spectroscopic features.

   \subsection{Stochastic subsampling and dynamic context} \label{Stochastic subsampling and dynamic context}
      Our input dataset comprises a vocabulary of approximately 120 distinct features (110 \textit{Gaia} XP coefficients plus 
      photometric and physical parameters). 
      To maintain computational efficiency and adhere to the model's context window limit of $N_{\text{ctx}}=64$, we implement a 
      stochastic subsampling strategy during training.

      For each training epoch, instead of feeding a fixed set of features, we randomly select $5$ to $64$ non-NaN tokens from the 
      available observations for each star to serve as the encoder input. 
      This strategy serves two critical purposes:
      \begin{enumerate}
         \item \textbf{Handling High Dimensionality:} It enables the model to learn from a total feature set that is larger than its 
         context window while naturally accommodating the missing data patterns inherent in astronomical surveys.
         
         \item \textbf{Breaking feature co-adaptation:} By randomly restricting the number of input tokens 
         (sometimes providing only a few), we force the model to break co-adaptations between specific data points. 
         This prevents the network from over-relying on dominant features (e.g., if parameter $C$ is predictable from both $A$ 
         and $B$, but $A$ is stronger, the model might ignore $B$ if $A$ is always present). 
         Consequently, the model learns robust relationships among all data types, significantly enhancing its generalization 
         capability.
      \end{enumerate}

   \subsection{Training and optimization} \label{Training and optimization}
      The model was implemented in PyTorch and trained on a single NVIDIA GeForce RTX 4060 GPU using mixed precision. 
      The optimization objective is to maximize the robust Gaussian Log-Likelihood (GLL) of the target values.

      To fully utilize the prior knowledge available in the dataset, our loss function explicitly accounts for the measurement 
      uncertainties in the ground truth. For a requested parameter with a ground-truth value $y$ and a measurement error 
      $\sigma_{\text{known}}$, if the model predicts a mean $\hat{y}$ and an internal predictive uncertainty 
      $\sigma_{\text{pred}}$, the loss function $J$ is defined as:
      \begin{equation}
         J(y, \hat{y}) = \frac{(y - \hat{y})^2}{2(\sigma_{\text{known}}^2 + \sigma_{\text{pred}}^2)} + \frac{1}{2} \ln(\sigma_{\text{known}}^2 + \sigma_{\text{pred}}^2)
      \end{equation}
      
      This formulation ensures that the model's predicted uncertainty ($\sigma_{\text{pred}}$) represents the intrinsic aleatoric 
      uncertainty after deconvolving the known measurement noise ($\sigma_{\text{known}}$).

      We trained the model for 4096 epochs with a batch size of 1024 using the AdamW optimizer. 
      The learning rate was managed via a Cosine Annealing Warm Restarts scheduler \citep{2016arXiv160803983L}, with a restart 
      period of $T_0 = 512$ epochs, decaying from an initial $\eta = 10^{-4}$ to a minimum of $\eta_{\text{min}} = 10^{-10}$. 
      To prevent overfitting, we applied a dropout rate of 0.1 in the dense layers and used Layer Normalization within the 
      Transformer blocks.

   \subsection{Inference and evaluation metrics} \label{Inference and evaluation metrics}
      During the inference stage, the model provides probabilistic outputs, yielding both a mean $\mu$ and a standard deviation 
      $\sigma$ for each requested parameter. 
      This standard deviation $\sigma$ quantifies the aleatoric uncertainty and serves as our primary metric for evaluating the 
      reliability of individual parameter predictions. 
      Since stellar age $\tau$ is trained in the logarithmic space ($\ln \tau$), we transform its predicted uncertainty back to 
      the linear space using a first-order Taylor expansion approximation: $\sigma_{\tau} \approx \tau \cdot \sigma_{\ln \tau}$.

      To validate the generative capability of the model (i.e., reconstructing spectra from physical parameters), we define the 
      spectral reconstruction error. 
      For a spectral sequence containing $M$ wavelength points (or coefficients), we calculate the Root Mean Square Error (RMSE) 
      between the generated spectra ($F_{\text{gen}}$) and the observed spectra ($F_{\text{obs}}$):

      \begin{equation}
         \mathrm{RMSE}_{\text{spec}} = \sqrt{ \frac{1}{M} \sum_{j=1}^{M} (F_{\text{gen}, j} - F_{\text{obs}, j})^2 }
         \label{eq:rmse_spec}
      \end{equation}

\section{Results} \label{sec:result}
   To demonstrate the general capabilities of our foundation model extended to evolved giant stars, we evaluate its 
   performance across multiple inference tasks following the experimental framework established by \cite{2024MNRAS.527.1494L}. 
   Unlike traditional machine-learning methods that typically require training independent models for specific parameters, 
   our model adopts the unified Transformer-based architecture proposed by \cite{2024MNRAS.527.1494L}, where all weights participate 
   during the inference process. 
   This allows for the simultaneous prediction of stellar atmospheric parameters ($T_{\text{eff}}$, $\log g$, $\text{[M/H]}$) 
   as well as key evolutionary parameters (mass and age) newly introduced via the APOGEE DR17 catalog.

   In the following subsections, we present the model's performance on various tasks: mapping stellar spectra directly to 
   stellar parameters (subsection \ref{sec:spectra_to_params}), 
   generating spectra from stellar parameters (subsection \ref{sec:params_to_spectra}), 
   inferring missing spectral portions (subsection \ref{sec:spectra_to_spectra}), 
   parameter-to-parameter inference (subsection \ref{sec:params_to_params}), 
   and recovering the interstellar extinction curve (subsection \ref{sec:extinction}).

   Unless otherwise specified, the scatter $\sigma$ refers to a robust measurement based on the median absolute deviation 
   (MAD), defined as $\sigma = 1.4826 \times \text{MAD}$. 
   This scaling factor ensures that for a Gaussian distribution, $\sigma$ is equivalent to the standard deviation. 
   All results presented are obtained by applying the model to an independent test set, which shares the same distribution as 
   the training data but was strictly excluded from the training process.

   \subsection{Spectra to stellar parameters} \label{sec:spectra_to_params}
      Deep learning provides a means to rapidly infer stellar parameters directly from stellar spectra. 
      As a primary test of our model, we evaluate its ability to infer both atmospheric and evolutionary parameters under different 
      input combinations and verify the rationality of its uncertainty estimation.
      \begin{figure*}
         \begin{center}
            \includegraphics[width=\linewidth]{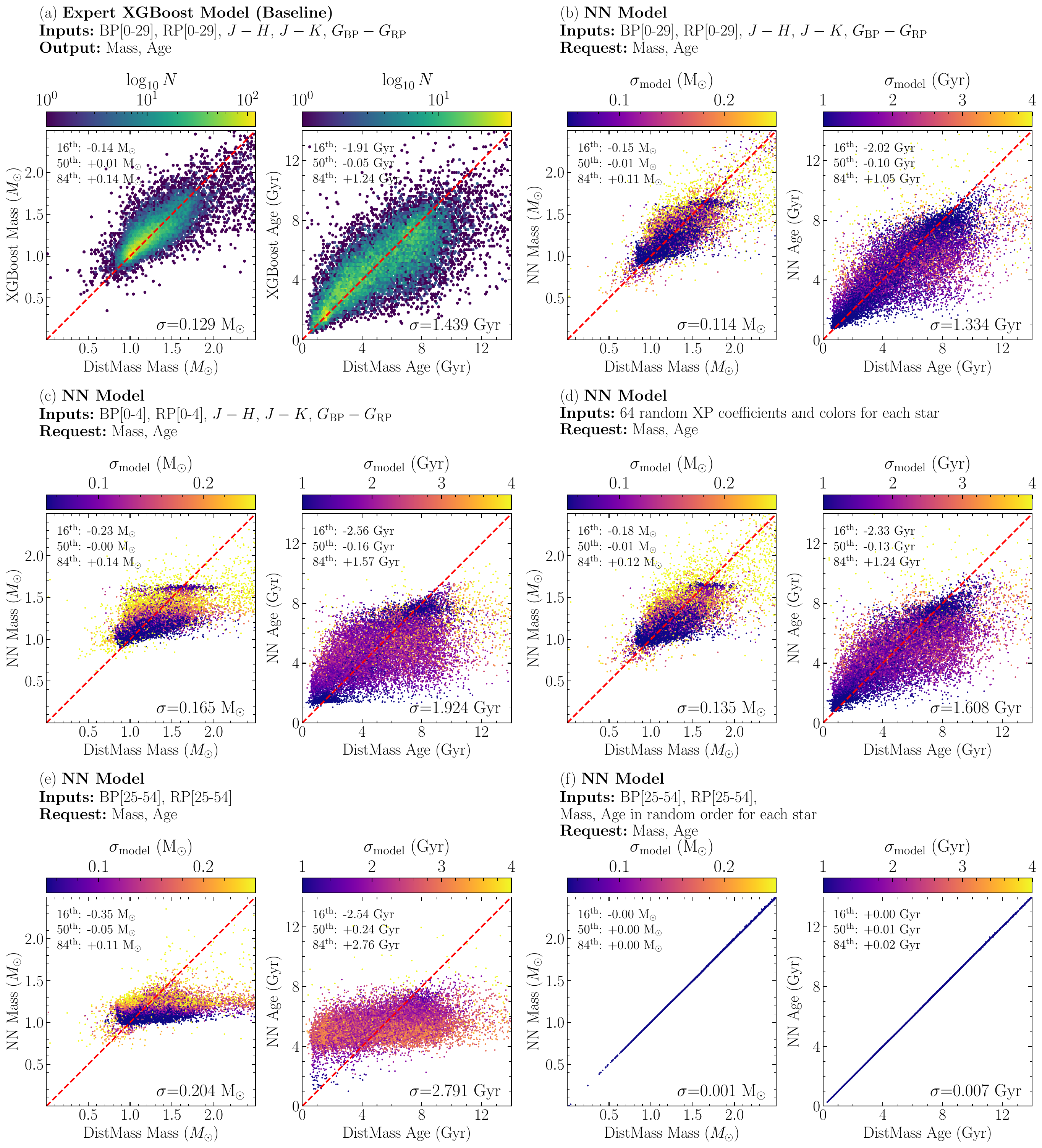}  
            \end{center}
            \caption{
               Results for predicting evolutionary parameters (mass and age). 
               \textit{(a)} The XGBoost baseline model performance. 
               \textit{(b)} Our Transformer-based model demonstrates superior performance, achieving 
               $\sigma_{\text{mass}} = 0.114\,M_{\odot}$ and $\sigma_{\text{age}} = 1.334\,\mathrm{Gyr}$. 
               Panels \textit{(c)} through \textit{(f)} show the model's robustness on mass and Age inference under varying 
               input conditions.}
            \label{fig:spectra_to_params_evol}
      \end{figure*}

      \begin{figure*}
         \begin{center}
            \includegraphics[width=\linewidth]{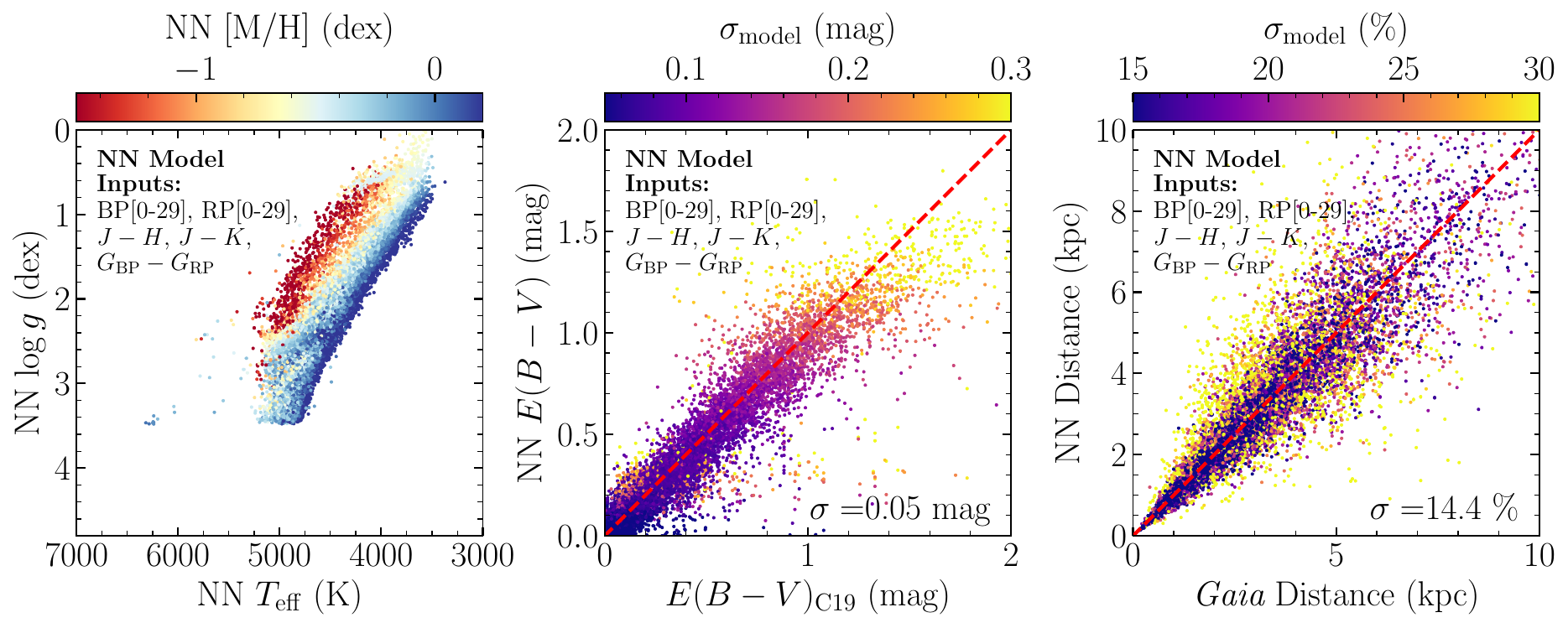}  
            \end{center}
            \caption{
               Physical validation of the model predictions. 
               \textit{Left:} The Kiel diagram ($T_{\text{eff}}$ vs. $\log g$) constructed from model-predicted parameters. 
               \textit{Middle:} Comparison between the model-inferred interstellar extinction $E(B-V)$ and reference values 
               ($\sigma=0.05\,\mathrm{mag}$). 
               \textit{Right:} Comparison between the model-inferred spectrophotometric distances and \textit{Gaia} geometric 
               distances ($\sigma=14.4\%$).}
         \label{fig:physical_validation}
      \end{figure*}

      \subsubsection{Atmospheric parameters and robustness tests}
         To place our results in a reasonable context, we adopt the same baseline comparison strategy as \cite{2024MNRAS.527.1494L}. 
         Specifically, we trained a dedicated XGBoost model using inputs identical to the standard configuration of our neural 
         network: the first 30 BP and RP coefficients, as well as $J-H$, $J-K$, and $G_{\mathrm{BP}}-G_{\mathrm{RP}}$ colors. 
         The comparison results are shown in Figure \ref{fig:spectra_to_params_atm}.

         Figure \ref{fig:spectra_to_params_atm}(a) displays the performance of the baseline XGBoost model, which achieves a scatter of 
         $\sigma = 51.73\,\mathrm{K}$ for $T_{\text{eff}}$, $0.16\,\mathrm{dex}$ for $\log g$, 
         and $0.09\,\mathrm{dex}$ for $\text{[M/H]}$. 
         In comparison, our Transformer-based model (Figure \ref{fig:spectra_to_params_atm}(b)) achieves higher precision with the 
         same inputs, reducing the scatter to $\sigma_{T_{\text{eff}}} = 41.96\,\mathrm{K}$, $\sigma_{\log g} = 0.13\,\mathrm{dex}$, 
         and $\sigma_{\text{[M/H]}} = 0.08\,\mathrm{dex}$. 
         This result reproduces and validates the findings of \cite{2024MNRAS.527.1494L}, indicating that the general-purpose model slightly 
         outperforms the specialized traditional model in fundamental parameter inference.

         Furthermore, the model demonstrates exceptional robustness to missing data. 
         As shown in Figure \ref{fig:spectra_to_params_atm}(c), when provided with only the first 5 BP/RP coefficients, the model 
         maintains an accuracy of $\sigma_{T_{\text{eff}}} \approx 56\,\mathrm{K}$. 
         Figure \ref{fig:spectra_to_params_atm}(d) further illustrates the model's ability to handle fragmented information: 
         it performs robustly even when the input consists of 64 randomly selected XP coefficients or colors. 
         Figure \ref{fig:spectra_to_params_atm}(e) and (f) demonstrate the self-consistency of the model. When given high-order 
         coefficients (indices 25--54) dominated by noise, the model correctly predicts large uncertainties 
         (indicated by yellow points). 
         Conversely, when ground truth labels are mixed into the input sequence, the model recovers this information perfectly 
         ($\sigma \approx 0$), confirming its capability to utilize all available information within the context window.
      \subsubsection{Evolutionary parameters}
         Building upon the successful validation of fundamental atmospheric parameters, this study further extends the inference 
         capability to stellar mass and age. 
         These two parameters are traditionally difficult to measure directly from spectra alone due to their highly non-linear 
         and degenerate relationships with spectral features.

         Figure \ref{fig:spectra_to_params_evol} presents the model's performance in inferring these evolutionary parameters, 
         again compared against the XGBoost baseline. 
         For stellar mass, the XGBoost baseline model (Figure \ref{fig:spectra_to_params_evol}(a)) yields a scatter of 
         $\sigma = 0.129\,M_{\odot}$, whereas our model (Figure \ref{fig:spectra_to_params_evol}(b)) achieves a higher precision 
         of $\sigma = 0.114\,M_{\odot}$. 
         Similarly, for stellar age inference, our model reduces the scatter from $1.439\,\mathrm{Gyr}$ (baseline) to 
         $1.334\,\mathrm{Gyr}$. 
         This significant performance improvement suggests that the Transformer-based architecture can more effectively capture the 
         subtle evolutionary features implicit in the spectra, enabling reliable estimation of mass and age for a large sample of 
         giant stars. 
         The robustness tests in Figure \ref{fig:spectra_to_params_evol}(c--f) further confirm that the model's predictions for mass 
         and age remain within a reasonable physical range even under constrained input conditions.

      \subsubsection{Physical validation}
         To verify the physical self-consistency of the model's predictions, Figure \ref{fig:physical_validation} displays 
         diagnostic plots generated based on the predicted parameters. 
         The left panel of Figure \ref{fig:physical_validation} shows the distribution of test set stars in the model-predicted 
         $T_{\text{eff}}$--$\log g$ plane (Kiel diagram). 
         The distribution clearly reproduces the RGB and RC features, and the metallicity distribution 
         (color-coded) aligns with stellar evolution theory without exhibiting unphysical artifacts.

         The middle panel of Figure \ref{fig:physical_validation} demonstrates the model's ability to recover interstellar extinction 
         $E(B-V)$, achieving a precision of $\sigma = 0.05\,\mathrm{mag}$ compared to external catalogs (e.g., C19/StarHorse). 
         The right panel shows the inference results for spectrophotometric distances; the distances inferred solely from spectral 
         and photometric information achieve an accuracy of $14.4\%$ when compared to \textit{Gaia} geometric distances. 
         These combined results indicate that our model has not merely fitted the labels but has genuinely learned the intrinsic 
         physical connections between stellar parameters and observational data.

   \subsection{Stellar parameters to spectra} \label{sec:params_to_spectra}
      Given the flexible input-output architecture of our Transformer-based model, we can invert the inference process described 
      in Section \ref{sec:spectra_to_params}. 
      Instead of predicting parameters from spectra, we can task the model with generating \textit{Gaia} XP spectra directly from a given 
      set of stellar parameters. 
      This capability is crucial for synthetic data generation, missing data imputation, and verifying the model's physical 
      understanding of stellar properties. 
      In this section, we evaluate the model's performance in generating spectra for evolved giant stars, specifically 
      incorporating the newly added evolutionary parameters: mass and Age.
      \begin{figure*}[htp!]
         \centering
         \includegraphics[width=1.0\textwidth]{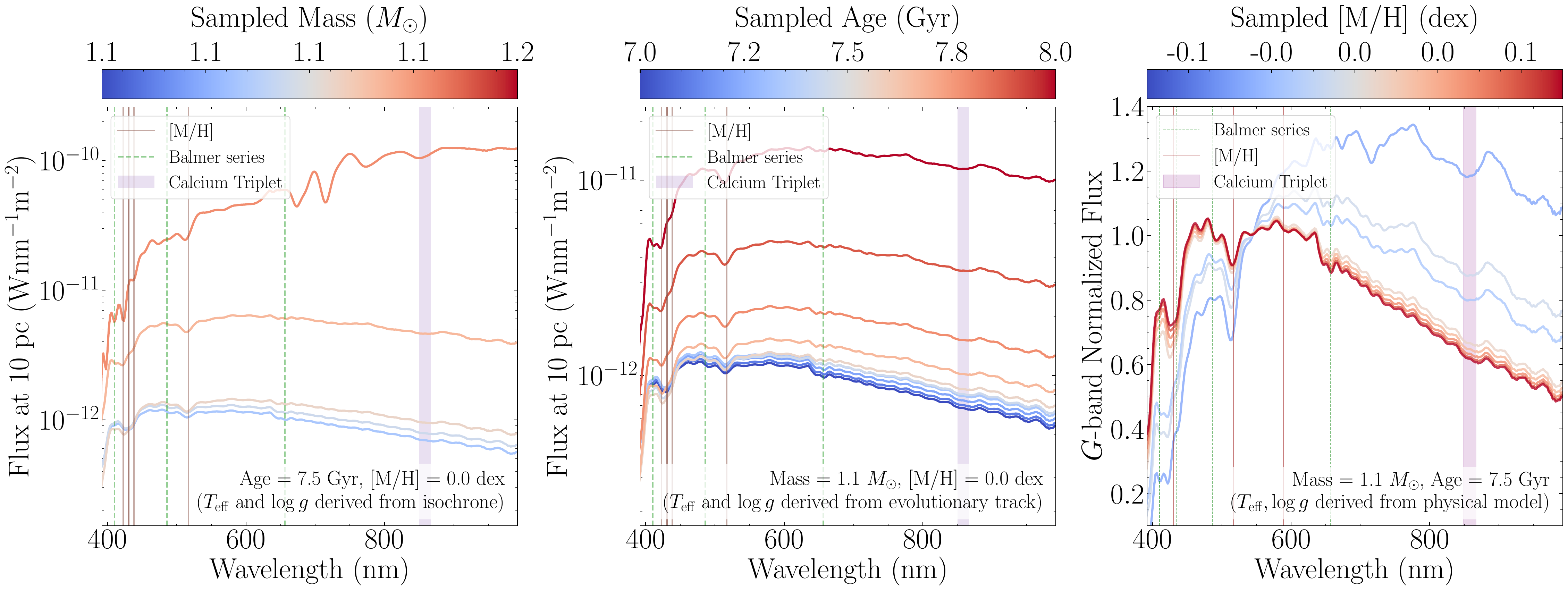}
         \caption{
            Model predicted spectra under physically consistent constraints. 
            The panels demonstrate the sensitivity of the model to key stellar parameters when constrained by MIST isochrones. 
            \textit{Left:} Generated spectra (flux at 10 pc) for stars with varying mass ($1.05 - 1.15 M_{\odot}$) at fixed Age 
            ($7.5$ Gyr) and Metallicity ($0.0$ dex). 
            The model correctly predicts higher luminosity for more massive giants. 
            \textit{Middle:} Spectra for stars with varying Age ($7.0 - 8.0$ Gyr) at fixed mass ($1.1 M_{\odot}$). 
            \textit{Right:} G-band normalized spectra showing the effect of [M/H] variations. 
            Deepening of metal absorption lines is clearly reproduced as [M/H] increases.}
         \label{fig:phys_consistency}
      \end{figure*}

      \begin{figure*}[htp!]
         \centering
         \includegraphics[width=1.0\textwidth]{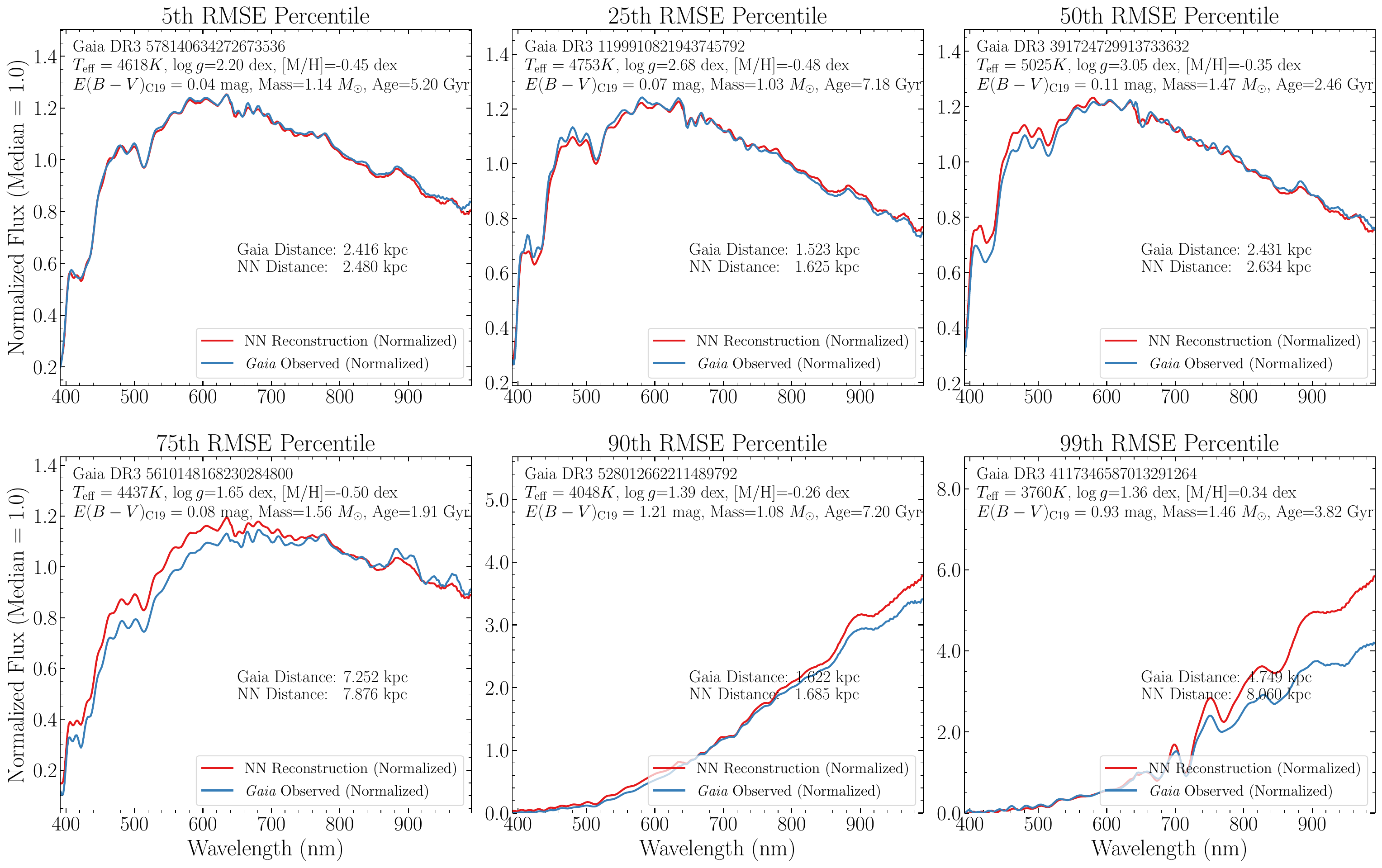}
         \caption{
            Comparison between observed and reconstructed \textit{Gaia} XP spectra. 
            Six examples are selected from the test set representing different percentiles of the reconstruction error (RMSE), 
            from the 5th (best) to the 99th (worst). 
            The blue lines represent the observed \textit{Gaia} spectra, and the red lines represent the spectra generated by our model 
            given the ground-truth parameters. 
            The model demonstrates high fidelity in reconstructing spectral shapes and features, even for cases with relatively 
            high reconstruction errors. 
            The inferred photometric distances (NN Distance) are also consistent with \textit{Gaia} geometric distances.}
         \label{fig:spec_percentiles}
      \end{figure*}

      \begin{figure*}[htp!]
         \centering
         \includegraphics[width=1.0\textwidth]{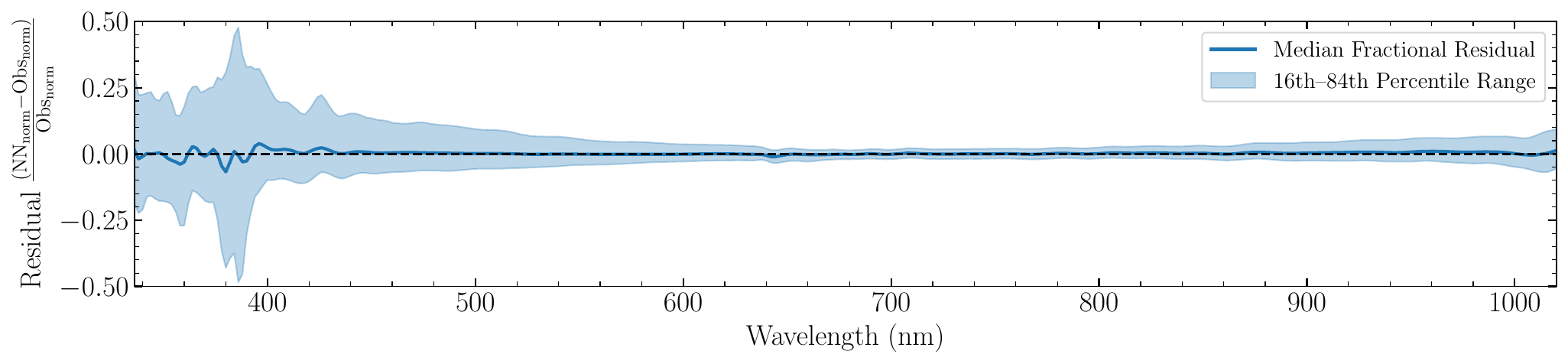}
         \caption{
            Wavelength-dependent fractional residuals.
            The solid blue line shows the median fractional residual 
            $(\mathrm{NN}_{\mathrm{norm}} - \mathrm{Obs}_{\mathrm{norm}}) / \mathrm{Obs}_{\mathrm{norm}}$ across the test set, 
            while the shaded region represents the 16th--84th percentile range. 
            The median residual remains close to zero across the full wavelength range, indicating no significant systematic 
            bias in the model's predictions. 
            The increased variance below 400 nm is attributed to the lower signal-to-noise ratio of \textit{Gaia} XP data in the blue channel.}
         \label{fig:wave_residuals}
      \end{figure*}

      \begin{figure}[!h]
      \begin{center}
       \includegraphics[width=\linewidth]{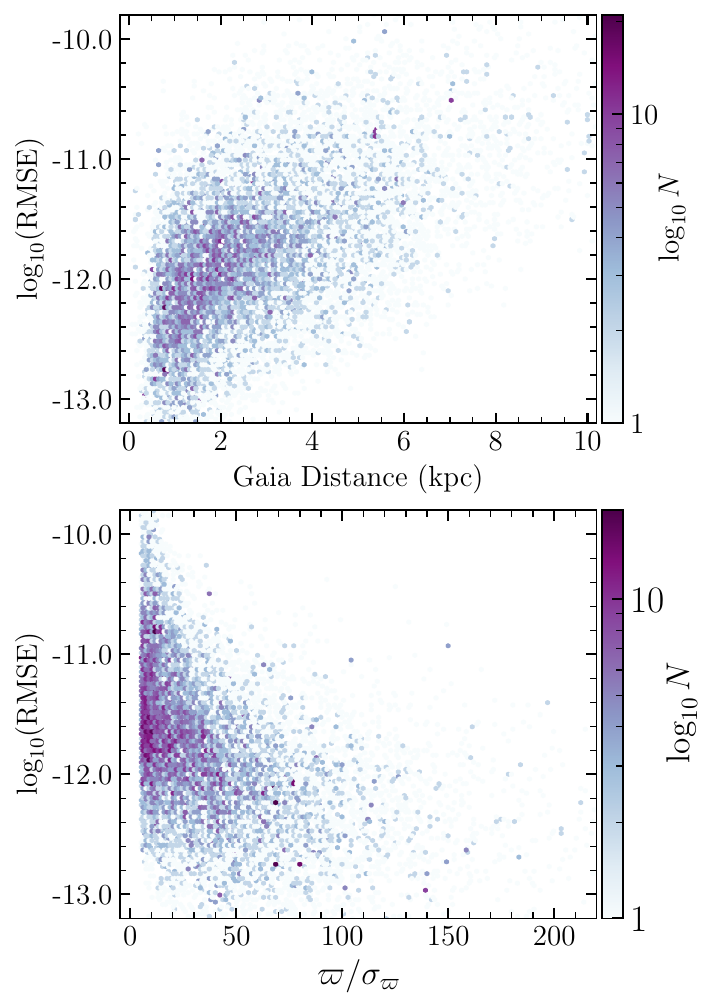}  
      \end{center}
      \caption{
         Diagnostic analysis of spectral reconstruction errors with respect to astrometric properties. 
         The panels show the distribution of $\log_{10}(\mathrm{RMSE})$ as a function of \textit{Gaia} distance (top) and Parallax 
         SNR $\varpi/\sigma_{\varpi}$ (bottom). 
         The reconstruction error is stable and low for nearby, high-SNR stars, and increases naturally with distance due to 
         observational noise inherent in the \textit{Gaia} XP spectra.}
         \label{fig:residual_dist_snr}
    \end{figure}

   \begin{figure*}[htp!]
      \centering
      \includegraphics[width=1.0\textwidth]{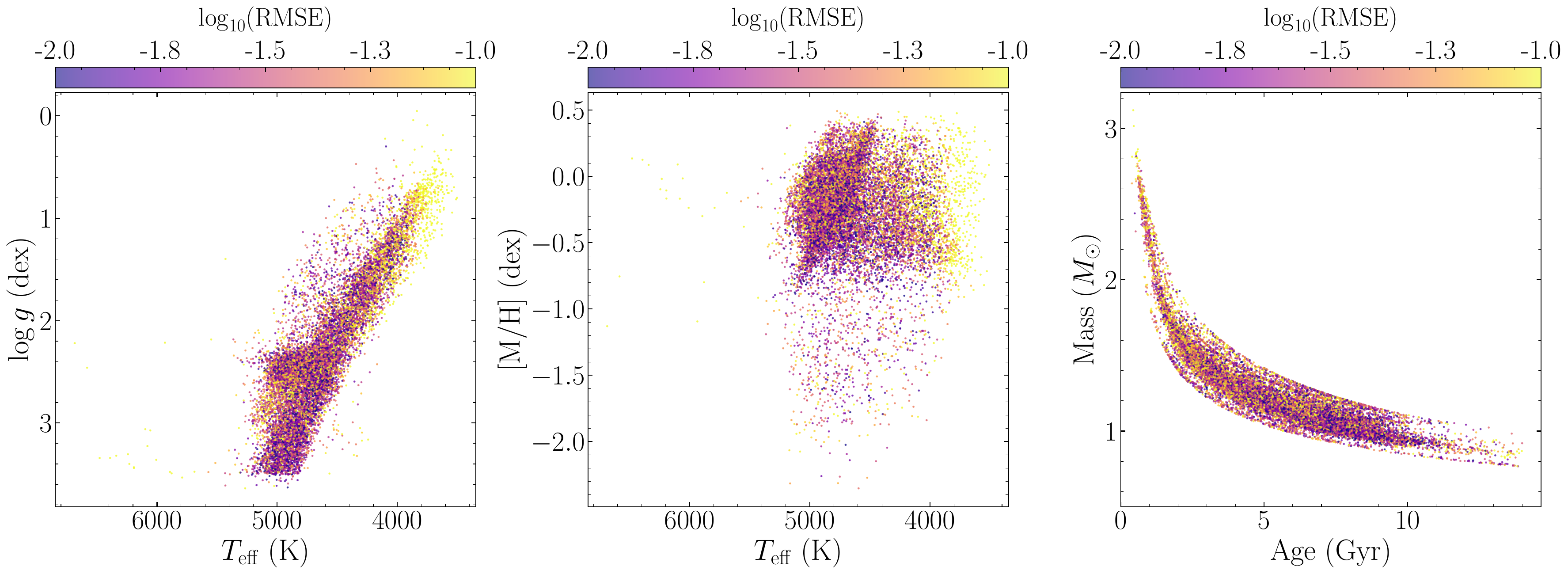}
      \caption{
         Distribution of reconstruction errors across the stellar parameter space. 
         The color-coded scatter plots display the RMSE distribution in the $T_{\text{eff}}$ vs. $\log g$ plane (left) and other 
         projections. 
         The consistent color distribution across the main locus of giant stars indicates that the model performs robustly across 
         the diverse population of evolved giants, without significant bias in specific evolutionary stages or [M/H] ranges.}
      \label{fig:residual_param_space}
   \end{figure*}

      \begin{figure*}[htp!]
         \centering
         \includegraphics[width=1.0\textwidth]{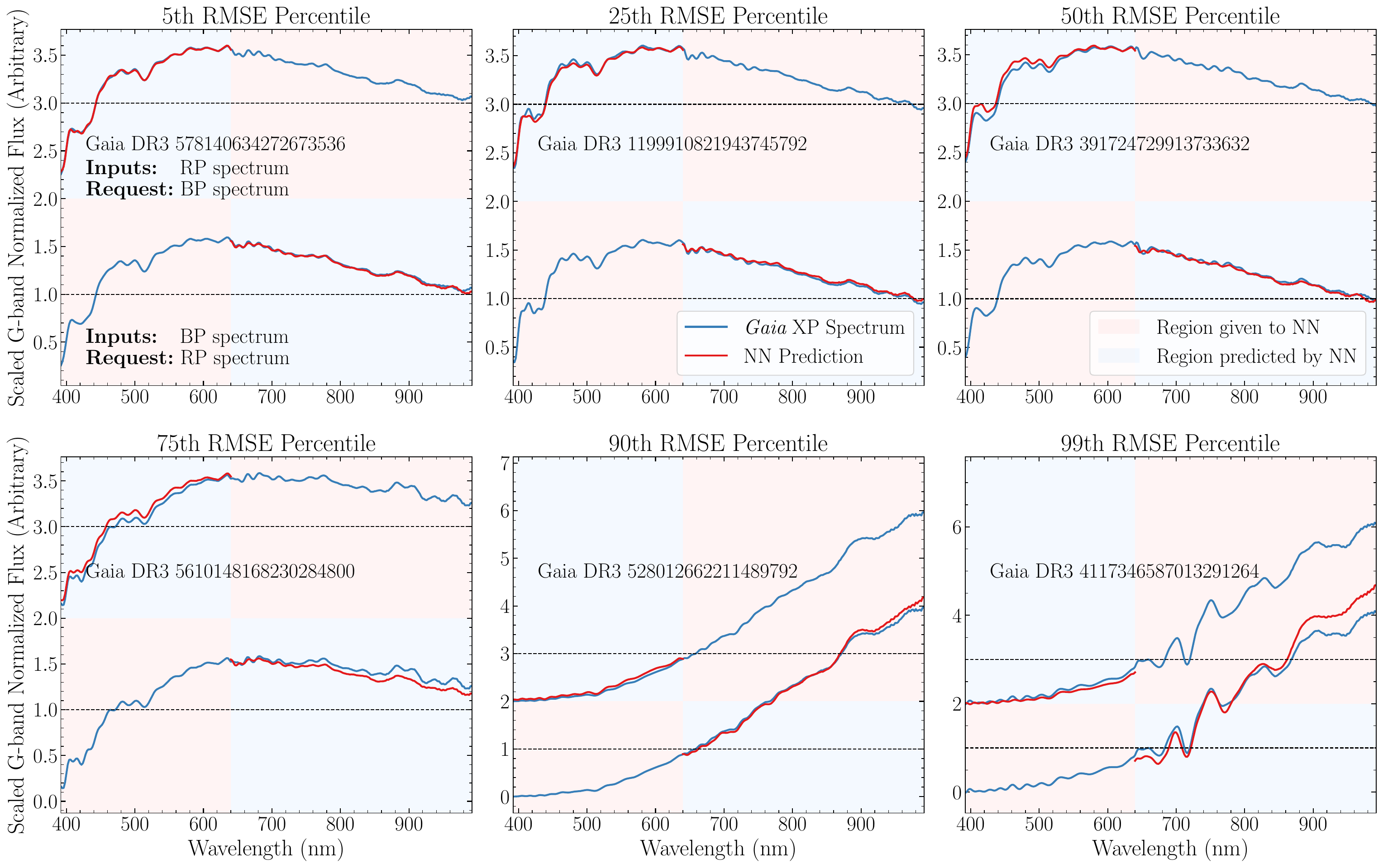}
         \caption{
            Spectra-to-spectra reconstruction based on partial spectral context. 
            The figure displays six representative stellar samples from the test set, corresponding to the 5th, 25th, 50th, 75th, 
            90th, and 99th percentiles of the reconstruction RMSE distribution, covering the range from best to worst fits. 
            Each panel is divided into upper and lower sections to demonstrate reciprocal inference tasks: 
            the upper section uses the red end of the spectrum (RP, light red background) as input to predict the blue end 
            (BP, light blue background); 
            the lower section uses the blue end (BP, light red background) as input to predict the red end (RP, light blue background). 
            The blue solid lines represent the observed \textit{Gaia} XP spectra, while the red solid lines indicate the neural 
            network predictions. 
            Black dashed lines are provided as reference levels for flux comparison. 
            The results show that even with minimal overlap between the input and predicted wavelengths, the model leverages its 
            deep understanding of stellar evolution physics to recover missing spectral features with high accuracy and robustness.}
         \label{fig:spec_recon}
      \end{figure*}

      \begin{figure*}[htp!]
         \centering
         \includegraphics[width=0.7\textwidth]{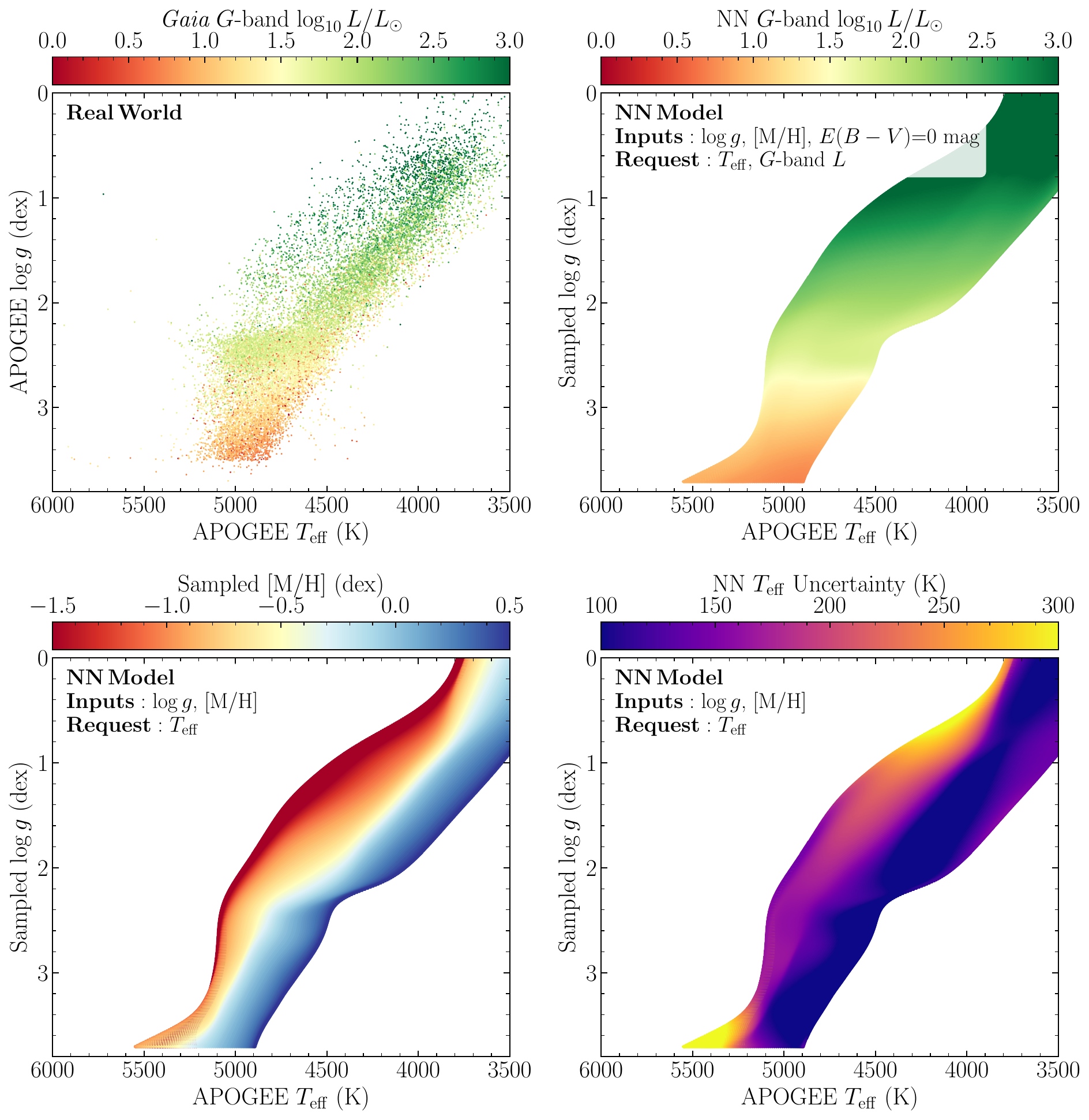}
         \caption{Verification of the physical manifold on the Kiel Diagram. 
            \textit{Top Left:} The distribution of real observational samples from APOGEE DR17, color-coded by the absolute 
            $G$-band luminosity ($\log_{10} L/L_{\odot}$) derived from \textit{Gaia} parallax. 
            \textit{Top Right:} Inference results of the model on an artificial uniform parameter grid 
            ($\log g \in [0, 4], [\mathrm{M/H}] \in [-2.5, 0.5]$). 
            \textit{Bottom Left:} The same model predictions as above, but color-coded by the input $[\mathrm{M/H}]$. 
            \textit{Bottom Right:} The distribution of uncertainty in the model's $T_{\text{eff}}$ predictions.}
            \label{fig:kiel_manifold}
      \end{figure*}

       \begin{figure*}[htp!]
            \centering
            \includegraphics[width=1.0\textwidth]{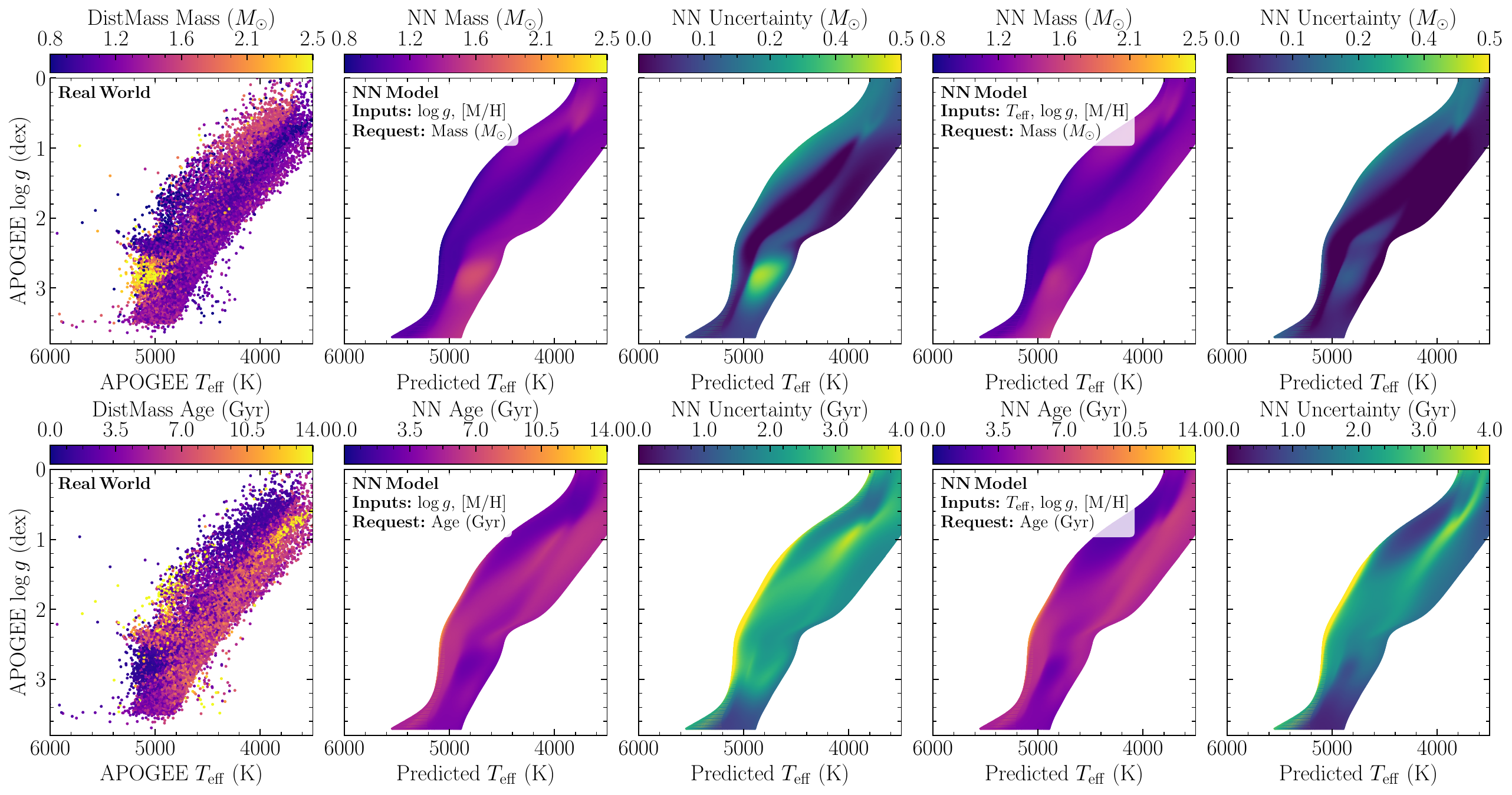}
            \caption{
               Self-consistent inference of evolutionary parameters (mass and Age). 
               The top row displays the inference results for mass, and the bottom row for Age. 
               \textit{Column 1:} Real data distribution from the APOGEE DR17 and DistMass Value-Added Catalog, serving as the 
               reference ground truth. 
               \textit{ 2--3 (NN Model):} The model receives only $\log g$ and $[\mathrm{M/H}]$ as input and must 
               simultaneously predict $T_{\text{eff}}$ ($x$-axis) and the target parameter (color). 
               \textit{Columns 4--5 (NN Model):} The $T_{\text{eff}}$ generated in \textit{Column 2} is fed back into the 
               model as a known condition.}
               \label{fig:mass_age_consistency}
         \end{figure*}
      \subsubsection{Physical consistency}\label{sec:phys_consistency}
         To verify that our model has genuinely learned the physical connections between stellar parameters and spectral 
         features—rather than merely performing interpolation or memorization of the training data—we conducted a series of 
         physically consistent experiments. 
         Instead of arbitrarily combining parameters, we queried MIST isochrones to ensure that the input tuples 
         ($T_{\text{eff}}$, $\log g$, mass, age) strictly adhere to stellar evolution theory.

         Figure \ref{fig:phys_consistency} demonstrates the spectral response of our model under these controlled physical 
         constraints. 
         In the left panel, we fixed the age ($7.5$ Gyr) and [M/H] ($0.0$ dex) while varying the stellar mass within a 
         narrow range ($1.05 M_{\odot}$ to $1.15 M_{\odot}$). Despite the subtle changes in mass, the model correctly predicts a 
         systematic variation in absolute flux (flux at 10 pc), with more massive stars exhibiting higher luminosity. 
         This indicates that the model has successfully captured the non-linear mass-luminosity relation specific to the giant branch.
         The middle panel illustrates the age effect: at a fixed mass ($1.1 M_{\odot}$), the model distinguishes the spectral 
         features corresponding to ages ranging from $7.0$ to $8.0$ Gyr. 
         The right panel highlights the effect of $\text{[M/H]}$ on the spectral shape. Using G-band normalized flux, 
         we observe a clear deepening of major absorption as $\text{[M/H]}$ increases, which is fully consistent 
         with stellar atmospheric theory. 
         These results strongly suggest that our Foundation Model has internalized the physics of stellar evolution.

      \subsubsection{Spectral reconstruction}\label{sec:spec_reconstruction}
         To quantitatively assess the accuracy of the generated spectra, we reconstructed the \textit{Gaia} XP spectra for all stars in the 
         test set using their ground-truth parameters and compared them with the actual observations.

         Figure \ref{fig:spec_percentiles} presents six representative examples selected based on their reconstruction RMSE, 
         ranging from the 5th percentile (best fit) to the 99th percentile (worst fit). 
         For the vast majority of the sample, including the 75th percentile, the reconstructed spectra (red solid lines) show 
         excellent agreement with the observed \textit{Gaia} spectra (blue solid lines), accurately reproducing both the continuum shape 
         and specific absorption features. Remarkably, even for the 99th percentile case—which typically corresponds to stars with 
         low signal-to-noise ratios (SNR) or extreme parameters—the model recovers the overall spectral profile with high robustness. 
         Furthermore, the "NN Distance" derived from the model-predicted absolute magnitude is highly consistent with the geometric 
         distance inferred from \textit{Gaia} parallax, further validating the accuracy of the predicted luminosity.

         The statistical properties of the residuals are shown in Figure \ref{fig:wave_residuals}. 
         The solid blue line represents the median fractional residual, defined as 
         $(\text{NN}_{\text{norm}} - \text{Obs}_{\text{norm}}) / \text{Obs}_{\text{norm}}$, across the entire test set. 
         The median remains close to zero throughout the wavelength range ($336-1020$ nm), indicating no significant systematic bias. 
         The shaded region (16th--84th percentile range) reveals that the reconstruction variance is higher in the blue end 
         ($< 400$ nm), which is attributed to the naturally lower SNR of \textit{Gaia} XP data in this channel and the stronger effects of 
         extinction. 
         In contrast, the reconstruction in the red and near-infrared bands is highly precise and stable.

      \subsubsection{Error distribution}\label{sec:error_analysis}
         Finally, we examine the distribution of reconstruction errors across different observational conditions and the stellar 
         parameter space.

         Figure \ref{fig:residual_dist_snr} displays the relationship between the reconstruction RMSE, \textit{Gaia} distance, and 
         parallax SNR ($\varpi/\sigma_{\varpi}$). 
         As expected, the RMSE shows a positive correlation with distance and a negative correlation with SNR. 
         For local stars ($d < 4$ kpc) or high-quality data ($\text{SNR} > 50$), the model achieves exceptional reconstruction 
         precision ($\log_{10}(\text{RMSE}) \approx -12$). 
         The slight degradation in performance at larger distances is dominated by observational noise rather than model limitations.

         Complementary to the astrometric dependence, Figure \ref{fig:residual_param_space} maps the RMSE onto the physical 
         parameter space defined by $T_{\text{eff}}$, $\log g$, and $\text{[M/H]}$. 
         The error distribution is largely uniform across the parameter space, with no distinct "failure modes." 
         A minor increase in error is observed only at the edges of the distribution 
         (e.g., extreme low $T_{\text{eff}}$ or low $\text{[M/H]}$) where training data is sparse. 
         Overall, this confirms that our model provides reliable spectral generation capabilities across the diverse population 
         of evolved giants.

   \subsection{Spectra to spectra} \label{sec:spectra_to_spectra}
      Beyond the mapping between parameter space and spectral space, our foundation model demonstrates powerful intra-modal 
      inference capabilities, enabling the recovery of missing spectral bands solely from local contextual information. 
      This capability indicates that the model has not merely memorized labels but has established a robust internal 
      representation of the stellar spectral manifold.

      To evaluate this performance, we designed a challenging ``spectral inpainting'' experiment. 
      We partition the \textit{Gaia} XP spectra into Blue Photometer (BP) and Red Photometer (RP) components using a cutoff at 640 nm. 
      Since the overlap between BP and RP is minimal, the input portion contains almost no direct flux information regarding the 
      missing portion. 
      We conduct two alternating tests: 
      (1) inputting the RP spectrum to predict the BP spectrum, 
      and (2) inputting the BP spectrum to predict the RP spectrum. 
      In this process, the model must implicitly infer the stellar physical state---including $T_{\text{eff}}$, $\log g$, and the 
      newly introduced evolutionary parameters (mass and age)---based on the features of the input fragment to accurately 
      reconstruct the remaining spectrum.

      Figure~\ref{fig:spec_recon} presents representative results from the test set across different reconstruction error (RMSE) 
      percentiles, ranging from the best fit (5th percentile) to the worst fit (99th percentile). 

      First, the results demonstrate high reconstruction precision. 
      For the vast majority of the sample (e.g., from the 5th to the 75th percentile), the predicted spectra (red solid lines) 
      show good agreement with the observed ground truth (blue solid lines) in both luminosity level and detailed absorption 
      features. 
      This proves that the model has internalized the physics of stellar evolution: although the input is only half of the spectrum, 
      it effectively constrains the stellar properties, allowing the model to derive the full-band information.

      Second, compared to the parameter-based generation discussed in Section \ref{sec:params_to_spectra}, the ``spectra-to-spectra''
      reconstruction presented here often yields higher fidelity in spectral details. 
      This is because the actual observed spectral fragments contain significantly richer physical constraints 
      (such as specific abundance patterns or extinction details) than a simplified set of parameters. 
      Notably, since our model incorporates mass and age during training, it exhibits enhanced discriminative power in handling 
      degeneracies between the RGB and the RC, successfully locking onto the correct evolutionary 
      track based on partial spectral features.

      Finally, even for the 99th percentile cases---which typically correspond to stars with low SNR or those located at the edges 
      of the parameter space---the model robustly recovers the correct continuum shape without generating unphysical artifacts or 
      oscillations. 
      This robustness further confirms that our model captures the intrinsic physical laws governing the spectral variations of 
      evolved giants.

   \subsection{Stellar parameters to stellar parameters} \label{sec:params_to_params}
      To assess whether our foundation model has genuinely internalized astrophysical knowledge---rather than merely mechanically 
      memorizing the mapping between spectra and parameters---we designed a series of ``Label-to-Label'' experiments. 
      In these tests, we masked the spectral information and specific input parameters, requiring the model to infer the complete 
      physical state of a star based only on partial physical conditions. 
      This approach essentially probes whether the low-dimensional manifold established by the model in its latent space conforms 
      to standard stellar evolution theory.
      \subsubsection{The Kiel diagram and the luminosity}
         First, we replicated the analysis of \citet{2024MNRAS.527.1494L} to test the model's grasp of the Hertzsprung-Russell (HR) 
         diagram morphology. 
         We constructed a uniform grid covering the parameter space of red giants 
         ($\log g \in [0, 4]$, $[\mathrm{M/H}] \in [-2.5, 0.5]$) and queried the model for the effective temperature 
         ($T_{\text{eff}}$) and luminosity at these grid points.

         As shown in Figure~\ref{fig:kiel_manifold}, despite the unphysical, artificial nature of the input grid, the 
         output  distribution (top right) successfully reproduces the morphology of the RGB observed in real data. 
         Even more impressively, the model accurately predicts $T_{\text{eff}}$ the absolute $G$-band luminosity based on the input parameters, 
         reproducing the physical law where luminosity rises exponentially as $\log g$ decreases. 
         Furthermore, the bottom-left panel of Figure~\ref{fig:kiel_manifold} clearly demonstrates the modulation of $T_{\text{eff}}$ 
         by $[\mathrm{M/H}]$: at a constant $\log g$, the model correctly predicts that metal-rich stars are cooler than their 
         metal-poor counterparts. 
         These results indicate that our model has effectively compressed and internalized the high-dimensional observational data 
         into a low-dimensional manifold that adheres to physical laws.

      \subsubsection{Consistency of mass and age}
         We further investigated the model's understanding of evolutionary parameters---mass and age. 
         These parameters are typically considered difficult to determine solely from basic atmospheric parameters due to strong 
         physical degeneracies on the HR diagram. 
         To verify whether the model has overcome this challenge, we compared two distinct inference modes 
         (as shown in Figure~\ref{fig:mass_age_consistency}).
         The model is fed only $\log g$ and $[\mathrm{M/H}]$ and is required to predict $T_{\text{eff}}$, mass, and age. 
         This simulates how the model constructs the evolutionary state of a star in the absence of $T_{\text{eff}}$ 
         information.
         The model is fed the complete set of $T_{\text{eff}}$, $\log g$, and $[\mathrm{M/H}]$, simulating a traditional spectral 
         parameter measurement task.

         The comparison in Figure~\ref{fig:mass_age_consistency} reveals a striking phenomenon: 
         the mean distributions of mass and age predictions in the generative mode are nearly identical to those in the 
         analytical mode (see the top row of the middle and right columns in Figure~\ref{fig:mass_age_consistency}).

         This result suggests that our foundation model has fully internalized the physical constraints governing red giant 
         evolution. 
         In the model's cognitive framework, once the $\log g$ and $[\mathrm{M/H}]$ are given, the star's evolutionary stage 
         (manifested as $T_{\text{eff}}$) and its intrinsic properties (mass and age) are strictly 
         confined to specific isochrone tracks. For instance, in the RC region ($\log g \approx 2.4$), even without 
         $T_{\text{eff}}$ input, the model accurately predicts that stars in this region possess higher mass characteristics 
         (top row, middle column of Figure~\ref{fig:mass_age_consistency}). 
         This proves that the model is not simply performing interpolation but has constructed a complete physical picture.

         Although the predicted means remain consistent, the uncertainty maps 
         (columns 3 and 5 of Figure~\ref{fig:mass_age_consistency}) exhibit significant differences. 
         When $T_{\text{eff}}$ is added as an explicit constraint, the prediction confidence across the entire parameter space 
         increases significantly (indicated by darker colors). 
         This demonstrates that while the introduction of $T_{\text{eff}}$ does not alter the model's fundamental judgment of 
         the physical evolutionary trajectory, it effectively compresses the probabilistic space of the prediction. 
         This reflects the model's robust Bayesian inference characteristics: epistemic uncertainty decreases as the information 
         content increases.

      \subsubsection{Emergent laws of Galactic chemical evolution}
         Beyond reproducing static stellar physical parameters, we further examined whether the model captured evolutionary 
         history at the Galactic scale. 
         It is worth emphasizing that our model was never explicitly taught the correlation between ``time'' and 
         ``chemical abundance'' during training; all such knowledge is implicitly encoded in the statistical distribution of 
         spectra and labels.

\begin{figure}[!h]
         \begin{center}
         \includegraphics[width=\linewidth]{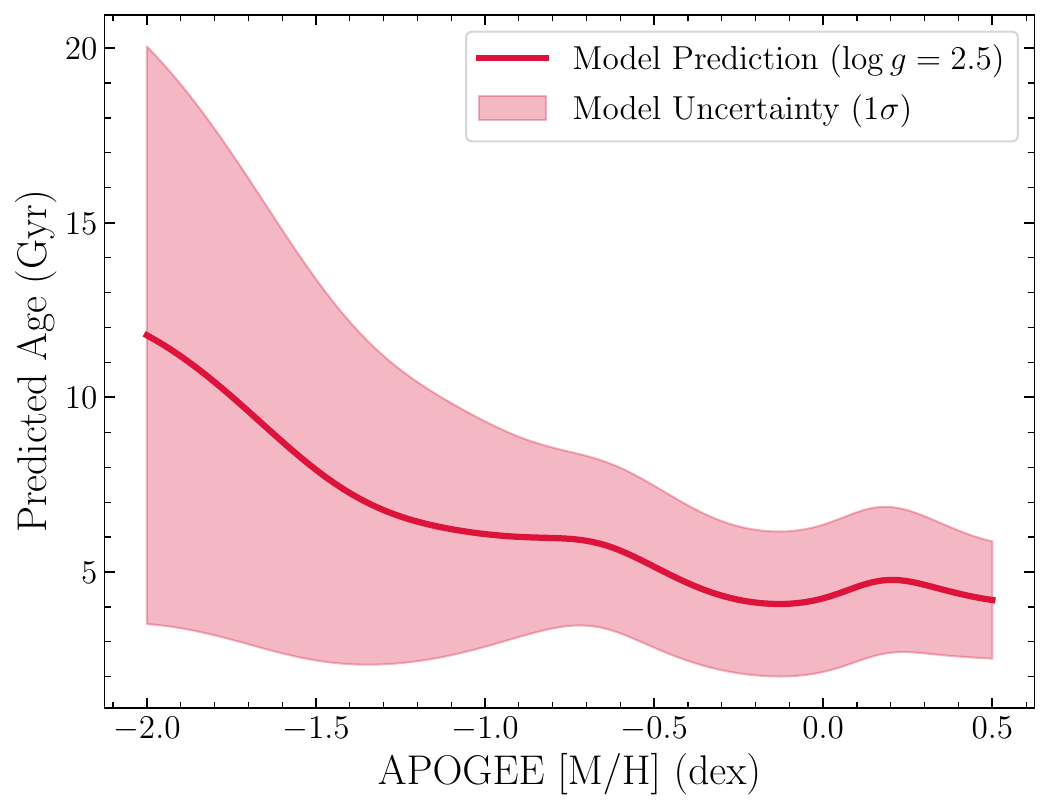}  
         \end{center}
         \caption{
            Emergent laws of Galactic chemical evolution. 
            The figure shows the relationship between stellar age and $[\mathrm{M/H}]$ as predicted by the model for a slice at 
            fixed surface gravity $\log g = 2.5$. 
            The solid line represents the predicted mean, and the shaded region indicates the $1\sigma$ uncertainty range.}
         \label{fig:age_metallicity}
      \end{figure}

         We designed a physical slice experiment: fixing the $\log g$ to a typical red giant value ($\log g = 2.5$), 
         we examined the relationship between the model-predicted age and $[\mathrm{M/H}]$. 
         As shown in Figure~\ref{fig:age_metallicity}, the model reveals a clear curve that aligns with astrophysical expectations:
         
         \begin{enumerate}
            \item \textbf{Age-Metallicity Relation (AMR):} 
            The model predicts an average age of approximately 12\,Gyr for metal-poor stars ($[\mathrm{M/H}] \approx -2.0$), 
            corresponding to the Galactic halo or thick disk components; conversely, stars near solar metallicity have an average 
            age of roughly 4--5\,Gyr, corresponding to the thin disk component.
            
            \item \textbf{Physically-Aware Uncertainty:} 
            The model's predicted uncertainty (shaded region in Figure~\ref{fig:age_metallicity}) increases significantly at the 
            metal-poor end. 
            This is not a performance defect but an accurate reflection of astrophysical reality---in ancient, metal-poor 
            isochrones, the evolutionary trajectories are highly crowded, where minute parameter perturbations lead to massive 
            age degeneracies.
         \end{enumerate}

         These results powerfully demonstrate that our foundation model is not merely a spectral analysis tool but a physics simulator. 
         It has successfully extracted and decoupled the physical laws governing stellar formation from the historical laws 
         governing Galactic evolution.

   \subsection{Recovery of the interstellar extinction curve} \label{sec:extinction}
      \begin{figure*}[htp!]
         \centering
         \includegraphics[width=1.0\textwidth]{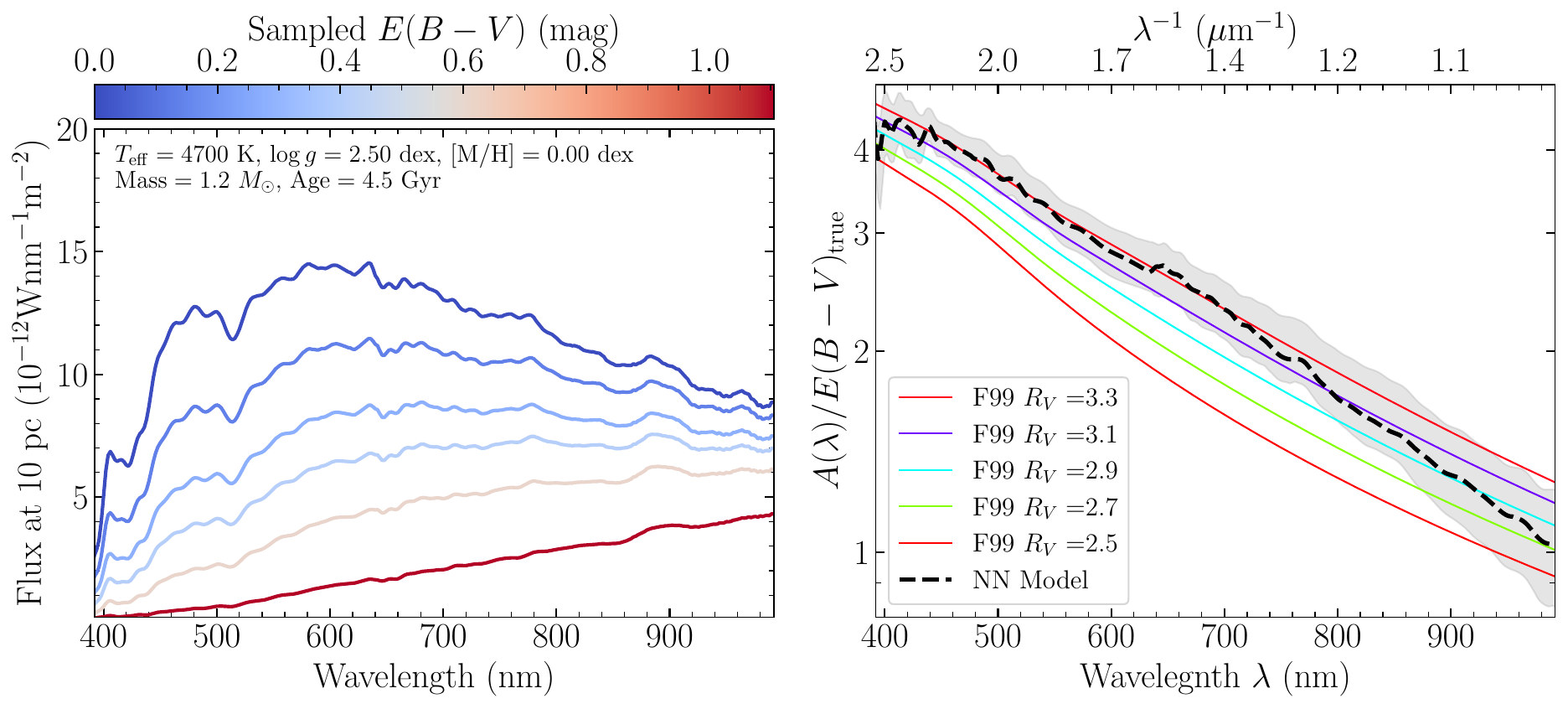}
         \caption{
            Recovery of the interstellar extinction curve by our foundation model. 
            \textit{Left:} The model's reconstruction of \textit{Gaia} XP spectra for a giant star with fixed intrinsic parameters 
            ($T_{\mathrm{eff}} = 4700$\,K, $\log g = 2.50$\,dex, $[\mathrm{M/H}]=0.0$\,dex, mass $= 1.2\,M_{\odot}$, and 
            age $= 4.5$\,Gyr) under varying reddening values $E(B-V)$. 
            \textit{Right:} The derived wavelength dependence of the extinction curve obtained by comparing the extinguished 
            spectra to the zero-extinction reference. 
            The black dashed line represents the median extinction curve $A(\lambda)/E(B-V)_{\mathrm{true}}$ learned by the model, 
            with the grey shaded region indicating the uncertainty (standard deviation) across the parameter grid. 
            The coloured solid lines show the standard extinction laws from \citet{1999PASP..111...63F} for different $R_V$ values. 
            The recovered curve shows strong agreement with the standard $R_V \approx 3.1$ law.}
         \label{fig:extinction_recovery}
      \end{figure*}

      As an example of an empirical law that is not explicitly included in the model formulation but can be recovered from its 
      latent representation, we predict the ratio of total-to-selective extinction, $R(\lambda)$, as a function of wavelength 
      $\lambda$. This ratio is defined as:

      \begin{equation}
         R(\lambda) = \frac{A(\lambda)}{E(B-V)_{\mathrm{true}}},
         \label{eq:extinction_ratio}
      \end{equation}
      where $A(\lambda)$ is the extinction at wavelength $\lambda$, and $E(B-V)_{\mathrm{true}}$ represents the reddening on 
      the true scale.

      Although our model incorporates no fixed or pre-defined extinction laws (unlike the hybrid approach of \cite{2023MNRAS.524.1855Z}), 
      it receives the $E(B-V)_{\mathrm{SFD}}$ value as a conditional input. 
      Consequently, the model must implicitly learn to disentangle the effects of interstellar extinction from the intrinsic stellar 
      spectral features. 
      To verify this capability, we derive the model's implied extinction curve by comparing simulated spectra of stars with 
      identical intrinsic physical parameters but varying degrees of reddening.

      We first established a grid of stellar parameters. 
      Since our work focuses on evolved giant stars, we fixed the metallicity to solar ($[\mathrm{M/H}]=0.0$) and varied the $\log g$ 
      from $1.0$ to $2.5$\,dex in steps of $0.1$\,dex. 
      The $T_{\mathrm{eff}}$ were determined using the empirical RGB relation for giants, defined as 
      $T_{\mathrm{eff}} = 5200 - 441.86\,(3.65 - \log g)$.
      Crucially, because our foundation model incorporates evolutionary parameters that were not present in the reference work, 
      we must ensure these are held constant to isolate the extinction effect. 
      Therefore, we fixed the stellar mass to a typical value of $1.2\,M_{\odot}$ and the age to $4.5\,\mathrm{Gyr}$ for all grid 
      points.

      For each parameter set in the grid, we emulated reference extinction-free spectra $F_0(\lambda)$ (where $E(B-V)=0$) and 
      extinguished spectra $F_E(\lambda)$ with input reddening values of $0.4$, $0.8$, and $1.2$\,mag. 
      The wavelength-dependent extinction curve is then calculated as follows:

      \begin{equation}
         R(\lambda) = \frac{A(\lambda)}{E(B-V)_{\mathrm{true}}} = \ln \left( \frac{F_0(\lambda)}{F_E(\lambda)} \right) 
         \frac{1}{0.884 \, E(B-V)_{\mathrm{SFD}}}.
         \label{eq:calc_extinction}
      \end{equation}

      We performed this calculation across the entire grid of stellar parameters and derived the median extinction curve. 
      The results are presented in Figure~\ref{fig:extinction_recovery}. 
      The left panel illustrates the spectral variations caused solely by extinction for a representative giant star. 
      The right panel displays the recovered extinction curve (black dashed line). 
      It is evident that our model's intrinsic extinction law is in excellent agreement with the standard mean extinction curve of 
      \cite{1999PASP..111...63F} with $R_V \approx 3.1$ (purple solid line). 
      We note that at the red end of the spectrum ($\lambda > 800$\,nm), the model prefers a slightly lower $R_V$, but the overall 
      consistency confirms that the model has successfully learned to physically decouple interstellar extinction from the 
      intrinsic stellar parameters (including mass and age) without relying on physical priors.

\section{Discussion}\label{sec:Discussion}
      \subsection{Physical interpretability of internal mechanisms}
         \begin{figure*}[htp!]
            \centering
            \includegraphics[width=1.0\textwidth]{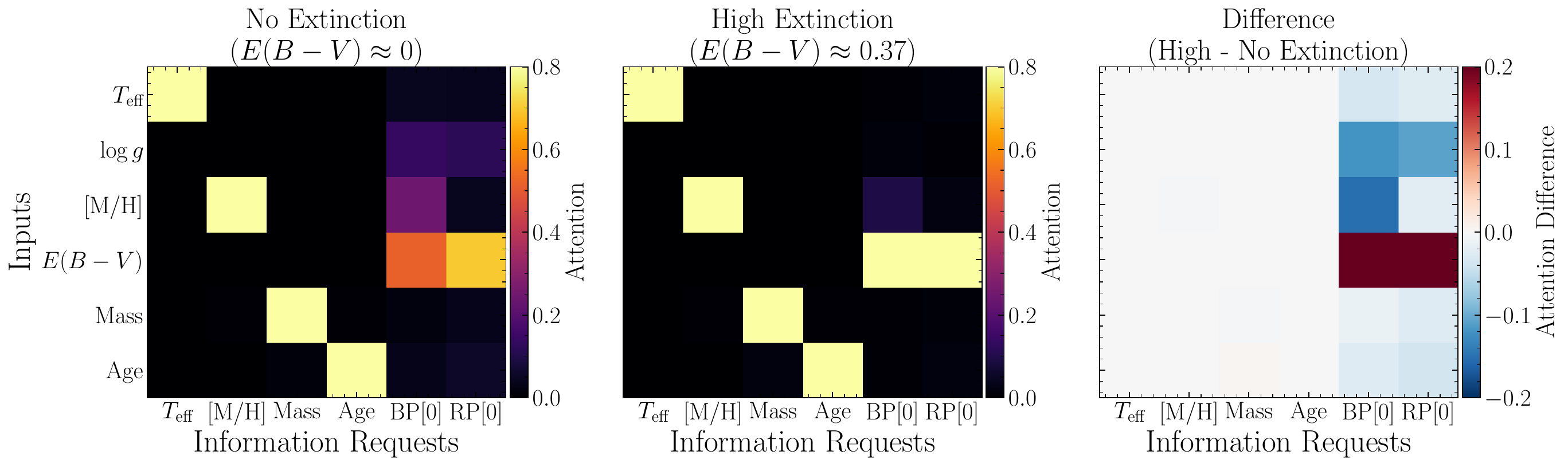}
            \caption{
                  Visualization of the cross-attention mechanism in the foundation model under different extinction scenarios. 
                  The maps display the attention scores between input parameters ($y$-axis) and requested outputs ($x$-axis). 
                  \textit{Left:} Attention weights for a star with negligible extinction ($E(B-V) \approx 0$). 
                  \textit{Middle:} Attention weights for the same star but with high extinction ($E(B-V) \approx 0.37$). 
                  \textit{Right:} The difference map (High Extinction $-$ No Extinction). 
                  Red regions indicate increased attention, while blue regions indicate decreased attention. 
                  Note that when requesting spectral coefficients (BP[0], RP[0]) under high extinction, the model significantly 
                  shifts its attention from $T_{\text{eff}}$ to $E(B-V)$, reflecting the physical dominance of reddening. 
                  Conversely, the attention for mass and age remains stable (white regions in the difference map), 
                  demonstrating the model's ability to disentangle intrinsic evolutionary parameters from extrinsic environmental 
                  effects.}
            \label{fig:attention_weights}
         \end{figure*}

         \begin{figure*}[htp!]
            \centering
            \includegraphics[width=1.0\textwidth]{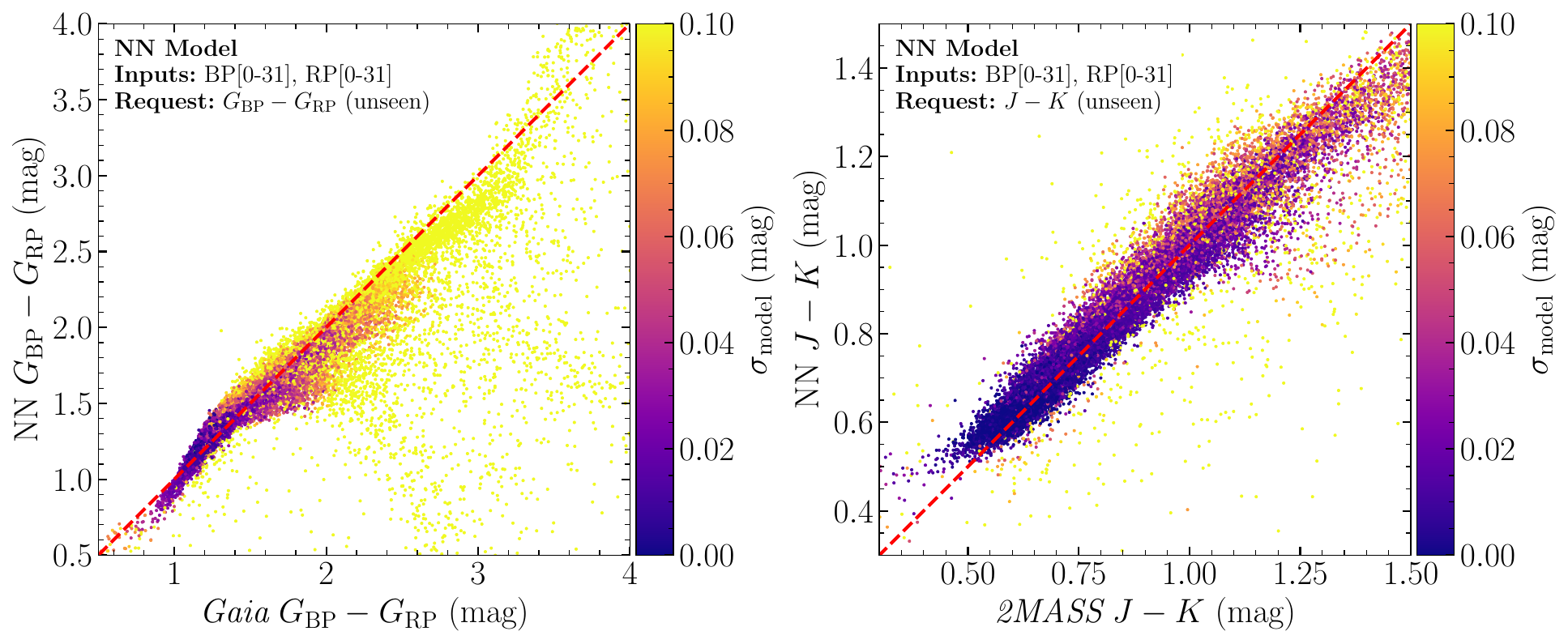}
            \caption{
               Evaluation of the physical meaningfulness of the learned embedding space using ``unseen'' labels. 
               The plots show the model's reconstruction of photometric colors (a) \textit{Gaia} $G_{\text{BP}} - G_{\text{RP}}$ 
               and (b) 2MASS $J-K$, based solely on the embedding derived from \textit{Gaia} XP spectral coefficients 
               ($\text{BP}[0\text{--}31]$, $\text{RP}[0\text{--}31]$). 
               The decoder was not explicitly optimized for these outputs during the primary training phase. 
               The color scale represents the model's predicted uncertainty ($\sigma_{\text{model}}$). 
               The tight correlation along the one-to-one line (red dashed) indicates that the embedding space effectively 
               encodes the full SED governed by stellar mass and age, allowing for accurate 
               generalization to derived observables.}
            \label{fig:photometric_colors}
         \end{figure*}

         \begin{figure*}[htp!]
            \centering
            \includegraphics[width=1.0\textwidth]{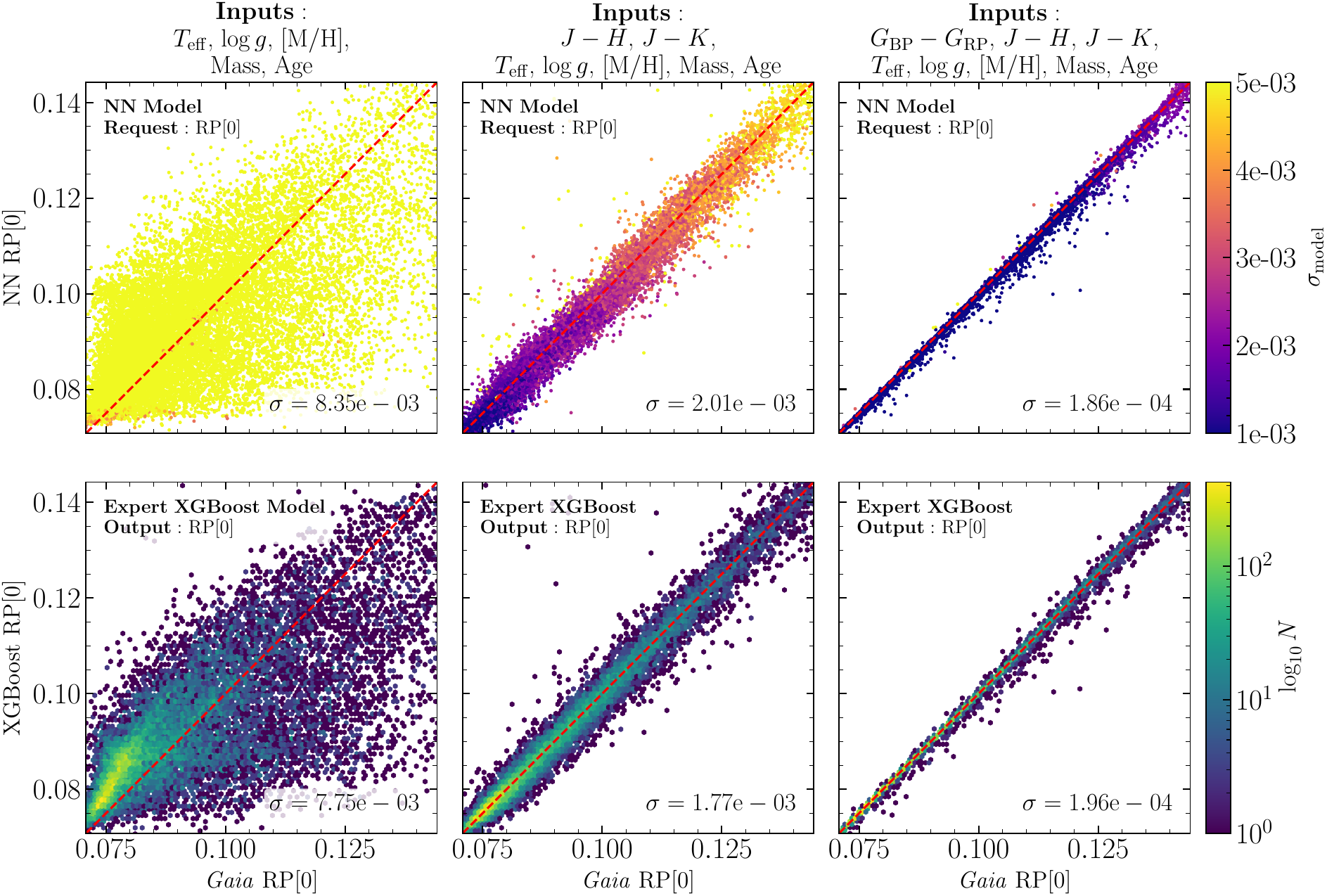}
            \caption{
               Comparison of the prediction performance for the first \textit{Gaia} RP spectral coefficient ($\text{RP}[0]$) between 
               our foundation model (\textit{top row}) and a specialized expert XGBoost model (\textit{bottom row}) 
               under varying input information. 
               \textit{Left column:} Inputs limited to stellar parameters ($T_{\text{eff}}$, $\log g$, $[\text{M/H}]$, mass, age) 
               without extinction info. 
               \textit{Middle column:} Inputs supplemented with near-infrared colors ($J-H$, $J-K$). 
               \textit{Right column:} Full inputs including optical color ($G_{\text{BP}} - G_{\text{RP}}$). 
               The top row is colored by the foundation model's predicted uncertainty ($\sigma_{\text{model}}$), while the bottom 
               row is colored by the logarithmic number density ($\log N$). 
               The scatter ($\sigma$) decreases significantly as color information is added, demonstrating that the foundation 
               model correctly quantifies the epistemic uncertainty caused by missing extinction data and matches the precision 
               of the expert model when full information is available.}
            \label{fig:model_comparison}
         \end{figure*}

         The interpretability of the transformer architecture allows us to verify whether the model learns physical laws or 
         merely memorizes statistical correlations. 
         Figure~\ref{fig:attention_weights} illustrates the cross-attention weights between input parameters and information requests 
         under two distinct scenarios: a ``No Extinction'' case ($E(B-V) \approx 0$) and a ``High Extinction'' case 
         ($E(B-V) \approx 0.37$). 
         The diagonal dominance in the attention matrices (left and middle panels) indicates that when the model is tasked with 
         reconstructing intrinsic labels---such as $T_{\text{eff}}$, $[\text{M/H}]$, mass, and age---it correctly identifies and 
         prioritizes the corresponding ground truth from the input sequence. 
         This confirms the model's ability to perform robust identity mapping for available stellar parameters.

         A more profound physical understanding is revealed when analyzing the model's reconstruction of the \textit{Gaia} XP spectral 
         coefficients (BP[0] and RP[0]), which govern the overall shape of the Spectral Energy Distribution (SED). 
         As shown in the difference map (right panel of Figure~\ref{fig:attention_weights}), there is a significant shift in attention 
         when extinction increases. 
         In the low-extinction regime, the model relies primarily on $T_{\text{eff}}$ to determine the spectral slope. 
         However, in the high-extinction regime, the attention on $T_{\text{eff}}$ decreases (represented by blue regions), while 
         the attention on $E(B-V)$ surges (represented by dark red regions). 
         This behavior demonstrates that the model has autonomously disentangled the effects of temperature and interstellar dust: 
         it recognizes that in heavily obscured regions, the observed spectral reddening is physically dominated by extinction 
         rather than the intrinsic $T_{\text{eff}}$.

         Crucially, the attention patterns for the newly introduced evolutionary parameters, mass and age, exhibit stability against 
         environmental variations. 
         In the difference map, the columns corresponding to mass and age requests are predominantly white (near-zero difference), 
         indicating that the model's attention mechanism for these parameters is virtually unaffected by changes in interstellar 
         extinction. 
         This implies that the foundation model successfully treats mass and age as intrinsic, immutable properties of the star, 
         distinct from external factors like reddening. 
         This disentanglement is a key prerequisite for reliable stellar population studies in the Galactic disk, where variable 
         extinction often confounds traditional parameter estimation.

      \subsection{Evolutionary meaning of the embedding space}
         To verify that our foundation model learns a physically meaningful representation of stars---rather than merely minimizing 
         a task-specific loss function---we evaluate its ability to predict observables that were not explicitly emphasized 
         during the optimization of the decoder. 
         Figure~\ref{fig:photometric_colors} demonstrates the model's capacity to reconstruct photometric colors, 
         $G_{\text{BP}} - G_{\text{RP}}$ and $J - K$, solely from the learned embedding of the input Gaia\textit{Gaia} XP coefficients 
         ($BP[0\text{--}31]$, $RP[0\text{--}31]$).

         The predictions exhibit a tight correlation along the one-to-one line for both optical and near-infrared colors. 
         Since the model was not trained to map spectra directly to these specific color indices in a supervised manner, this 
         success implies that the embedding space effectively encodes the full SED of the stars. 
         Given that the morphology of the SED is fundamentally governed by the star's evolutionary state---defined by its mass and 
         age---the accurate recovery of these ``unseen'' colors serves as indirect evidence that the embedding space is organized 
         according to true evolutionary tracks.

         Specifically, the robust prediction of $G_{\text{BP}} - G_{\text{RP}}$ confirms that the embedding captures Wien's 
         displacement law and effective temperature with high fidelity.
         More importantly, the accurate recovery of the near-infrared color $J - K$ indicates that the embedding maintains physical 
         continuity into the Rayleigh-Jeans tail of the spectrum. 
         This generalization capability suggests that by incorporating mass and age into the training set, we have regularized the 
         latent space, forcing the model to cluster stars based on their intrinsic evolutionary stages 
         (e.g., distinguishing Red Clump stars from RGB stars based on their SED nuances) rather than simple spectral similarities.

      \subsection{Reliability and comparison with other methods}
         We assess the reliability of our foundation model by comparing its performance against a specialized XGBoost regressor 
         (``Expert Model'') and discussing its advantages over traditional parameter estimation methods such as \textit{The Cannon} 
         \citep{2015ApJ...808...16N} or SPT \citep{2024A&A...683A.163Z}. 
         Figure~\ref{fig:model_comparison} evaluates the prediction of the first \textit{Gaia} RP spectral coefficient (RP[0]) 
         under varying input information scenarios.

         In the scenario where only stellar parameters ($T_{\text{eff}}$, $\log g$, $[\text{M/H}]$, mass, age) are provided 
         (left column), both our foundation model and the expert model exhibit a significant scatter 
         ($\sigma \approx 8.35 \times 10^{-3}$). 
         This large uncertainty is physically justified: without explicit extinction information (e.g., $E(B-V)$) or photometric 
         colors, the spectral shape is inherently degenerate. 
         Crucially, our model correctly quantifies this ambiguity by outputting larger predicted uncertainties 
         (indicated by lighter colors in the scatter plot), thereby avoiding the ``hallucinations'' often seen in generative 
         models when data is incomplete. 
         This ``honesty'' in uncertainty estimation is a significant improvement over deterministic methods like 
         \textit{The Cannon} or StarNet \citep{2021ApJ...906..130O}, which typically require fixed input vectors and may produce biased point estimates 
         when key spectral features are missing or masked.

         As additional information is introduced, the model demonstrates rapid convergence. 
         The inclusion of near-infrared colors ($J-H$, $J-K$) reduces the scatter by a factor of four 
         ($\sigma \approx 2.01 \times 10^{-3}$, middle column), confirming that the model can implicitly infer extinction to 
         correct optical spectral predictions. 
         When the full photometric information ($G_{\text{BP}} - G_{\text{RP}}$) is available (right column), the precision 
         reaches $\sigma \approx 1.86 \times 10^{-4}$, matching the expert XGBoost model.

         Unlike discriminative models such as \texttt{astroNN} \citep{2019MNRAS.483.3255L} or SPT, which are primarily optimized 
         to map fixed spectral vectors to stellar labels, our foundation model leverages a generative self-attention mechanism to aggregate 
         information from arbitrary subsets of the spectral sequence. 
         This allows for robust recovery of parameters even in scenarios where specific spectral features are missing or inputs 
         are incomplete. 
         While specialized regressors often treat the spectrum as a rigid input, our approach autonomously weights the available 
         information---whether it be partial spectral coefficients or photometric colors---to minimize the aleatoric uncertainty 
         derived from finite SNR of the spectra, and the epistemic uncertainty propagated from the ``ground truth'' labels 
         themselves, which are model-dependent values derived from isochrones rather than direct observables.

\section{Conclusion} \label{sec:conclusion}
   Building upon the architecture proposed by \cite{2024MNRAS.527.1494L}, we successfully extended the transformer-based astronomical 
   foundation model to the domain of evolved giant stars, integrating complex evolutionary parameters---mass and age---into a unified 
   inference framework. 
   By cross-matching \textit{Gaia} XP spectra with the APOGEE DR17 DistMass catalog, we demonstrated that a single, pre-trained model can 
   simultaneously predict atmospheric parameters ($T_{\text{eff}}$, $\log g$, $[\text{M/H}]$) and evolutionary states with high precision. 
   Specifically, our model achieves a prediction scatter of $\sigma \approx 0.114 \, M_{\odot}$ for stellar mass and 
   $\sigma \approx 1.334 \, \text{Gyr}$ for age on an independent test set, outperforming specialized XGBoost baselines while maintaining 
   robust performance across diverse signal-to-noise regimes.

   Beyond quantitative precision, our extensive analysis reveals that the foundation model has autonomously internalized fundamental 
   astrophysical laws rather than merely memorizing statistical correlations. 
   Through generative experiments, we verified that the model correctly reproduces the non-linear mass-luminosity relation and the 
   age-dependent spectral variations predicted by stellar evolution theory. 
   Furthermore, the analysis of attention mechanisms confirms that the model successfully disentangles intrinsic stellar properties 
   from environmental effects; it can dynamically shift its focus between temperature-sensitive and extinction-dominated spectral features, 
   recovering the standard interstellar extinction law without explicit physical priors.

   Methodologically, this work highlights the superiority of the generative Transformer architecture over traditional discriminative 
   approaches (e.g., \textit{The Cannon} or \texttt{astroNN}) and rigid regression models. 
   Unlike methods that require fixed input vectors, our model treats stellar data as flexible sequences, allowing for the robust recovery of 
   missing spectral bands (``inpainting'') and the synthesis of consistent physical parameters from partial observations. 
   This probabilistic nature also ensures ``honest'' uncertainty estimation, avoiding the overconfident predictions often plagued by 
   deterministic models when extrapolating to the edges of the parameter space.

   Looking ahead, this foundation model framework offers a powerful tool for Galactic archaeology, particularly for disentangling the complex 
   history of the Milky Way's disk using large samples of giant stars. 
   Future work will focus on expanding the training set to include multi-band photometry and time-domain asteroseismic labels, further 
   refining the model's ability to resolve the ages of the oldest stars. 
   As we enter the era of massive spectroscopic surveys, such physically aware, data-driven models will be essential for translating 
   raw photons into a coherent narrative of stellar and Galactic evolution.

\begin{acknowledgements} 
This study is supported by the Natural Science Foundation of Shandong Province under grant No. ZR2024MA063, 
No. ZR2022MA076, and No. ZR2022MA089, the science research grants from the China Manned Space Project with 
No. CMS-CSST-2021-B05 and CMS-CSST-2021-A08, and the National Natural Science Foundation of China (NSFC) under grants 
No. 11873037.
\end{acknowledgements}

\bibliographystyle{aa} 
\bibliography{sample631} 
\end{document}